\documentclass[aps,prl,twocolumn,superscriptaddress,export,longbibliography]{revtex4-2} 
\usepackage{xpatch}
\usepackage{hyperref}
\usepackage{microtype}
\usepackage[utf8]{inputenc}
\usepackage[german, english]{babel}
\usepackage[dvipsnames]{xcolor}
\usepackage{amsmath}

\usepackage[export]{adjustbox}
\usepackage{titlesec}
\usepackage{amssymb}
\usepackage{braket}
\usepackage{dsfont}
\usepackage{comment}

\usepackage{subfigure}
\usepackage{newtxtext}
\usepackage[smallerops]{newtxmath}

\DeclareMathAlphabet{\mathcal}{OMS}{cmsy}{m}{n}
\DeclareMathAlphabet{\mathsf}{OT1}{cmss}{m}{n}

\hypersetup{
    colorlinks = true,
    linkcolor  = blue,
    citecolor  = blue,
    urlcolor   = blue,
    linktocpage= true
}
\newcommand{\CZ}{\mathsf{CZ}}

\newcommand{\id}{\mathds{1}}
\renewcommand{\vec}[1]{\mathbf{#1}}

\def\Z#1{\mathbb{Z}_{#1}}
\def\U#1{{U}(#1)}

\makeatletter
\g@addto@macro\bfseries{\boldmath}
\makeatother

\makeatletter
\patchcmd{\@ssect@ltx}
    {\addcontentsline{toc}{#1}{\protect\numberline{}#8}}
    {}
    {}
    {}
\makeatother

\DeclareMathOperator{\Tr}{Tr}
\DeclareMathOperator{\Div}{div}

\begin{document}

\title{Ergodicity breaking provably robust to arbitrary perturbations}

\date{October 5, 2022}

\author{David T. Stephen}
\affiliation{Department of Physics and Center for Theory of Quantum Matter, University of Colorado Boulder, Boulder, Colorado 80309 USA}
\affiliation{Department of Physics, California Institute of Technology, Pasadena, California 91125, USA}
\author{Oliver Hart}
\affiliation{Department of Physics and Center for Theory of Quantum Matter, University of Colorado Boulder, Boulder, Colorado 80309 USA}
\author{Rahul M. Nandkishore}
\affiliation{Department of Physics and Center for Theory of Quantum Matter, University of Colorado Boulder, Boulder, Colorado 80309 USA}

\begin{abstract}
We present a new route to ergodicity breaking via Hilbert space fragmentation that displays an unprecedented level of robustness. Our construction relies on a single emergent (prethermal) conservation law. In the limit when the conservation law is exact, we prove the emergence of Hilbert space fragmentation with
an exponential number of frozen configurations.
We further prove that every frozen configuration is absolutely stable to arbitrary perturbations, to all finite orders in perturbation theory. In particular, our proof is not limited to symmetric perturbations, or to perturbations with compact support, but also applies to perturbations with long-range tails, and even to arbitrary geometrically nonlocal $k$-body perturbations, as long as $k/L \rightarrow 0$ in the thermodynamic limit, where $L$ is linear system size. Additionally, we identify one-form $\U1$ charges characterizing some non-frozen sectors, and discuss the dynamics starting from typical initial conditions, which we argue is best interpreted in terms of the magnetohydrodynamics of the emergent one-form symmetry.
\end{abstract}

\maketitle

When do quantum many-body systems break ergodicity, and fail to reach thermal equilibrium under their own dynamics? `Traditional' answers have included integrable~\cite{Baxter} and many-body localized \cite{mblarcmp, mblrmp} systems, both of which have extensively many conserved quantities. A more recent answer involves many-body scars \cite{shiraishimori, aklt, turner, scarsarcmp}, whereby typical initial conditions thermalize, but there exist special (low-entanglement) initial conditions that do not. More recently still, it was observed that the interplay of (finitely many) conservation laws can break ergodicity \cite{ppn}, a phenomenon that was later understood as arising from Hilbert space fragmentation (aka shattering) \cite{KHN, Sala}, whereby the unitary time evolution matrix block diagonalizes into exponentially many subsectors, with the dynamics unable to connect different subsectors \cite{Moudgalya, SLIOM, Yang2020confinement, chamon, commutant, YoshinagaIsing, hn}.

An important open question involves how {\it robust} ergodicity breaking is to perturbations. For integrable systems, and most systems hosting scars, it is not known if there is any class of perturbations to which the phenomenon is robust. Many-body localization has a proof of robustness \cite{imbrie}, but the proof is subtle, only works for short-range interacting systems (with at most exponential tails) in one spatial dimension, and even there has recently been called into question \cite{selspolkovnikov}. In contrast, the best-studied route to Hilbert space fragmentation (charge and dipole conservation) has a simple proof of robustness \cite{KHN}, which applies in arbitrary dimensions, but only to symmetry-respecting perturbations with bounded spatial range $\ell$. It is also known, however \cite{KHN, YoshinagaIsing, hn}, that if conservation laws are implemented {\it emergently}, as prethermal conservation laws \cite{ADHH}, then the requirement that perturbations respect the corresponding symmetries gets lifted. Thus, it is known how to obtain (prethermal) Hilbert space fragmentation that is robust to {\it arbitrary} perturbations with bounded spatial range $\ell$. The bounded spatial range, however, has been hitherto essential, and implies that the proofs of fragmentation do not work with, e.g., interactions that have long-range tails, not even ones that decay exponentially with distance.

In this work, we present a new route to ergodicity breaking via Hilbert space fragmentation which is \emph{provably} robust to \emph{any} perturbations, \emph{without} the requirement that perturbations have bounded spatial range. Similarly to Refs.~\cite{KHN, Sala, YoshinagaIsing, hn}, we obtain exponentially many `frozen' configurations, and similarly to Refs.~\cite{KHN, YoshinagaIsing, hn} we rely on prethermal (i.e., `emergent') implementation of conservation laws. However, unlike those prior works, {\it all} our frozen configurations are absolutely stable \cite{Keyserlingk} to arbitrary perturbations, without any restriction on the range of the perturbation. In particular, our proofs apply also to systems with interactions that have physically realistic long-range tails -- not just tails decaying exponentially with distance, but also tails that decay as power-law functions of distance. Indeed, our proofs even apply to fully geometrically nonlocal perturbations, as long as the perturbations are $k$-body (i.e., act on no more than $k$ qubits) with $k/L \rightarrow 0$ in the thermodynamic limit, where $L$ is linear system size. As such, our construction produces ergodicity breaking with an unprecedented degree of robustness, and opens a new direction for the study of non-ergodic quantum dynamics. The phenomenon moreover arises in a system with a `generalized Rydberg' interaction, which could plausibly be accessed in near-term experiments with synthetic quantum matter.

\begin{figure*}
    \centering
    \subfigure[]{{\label{fig:duality}\includegraphics[scale=0.3135]{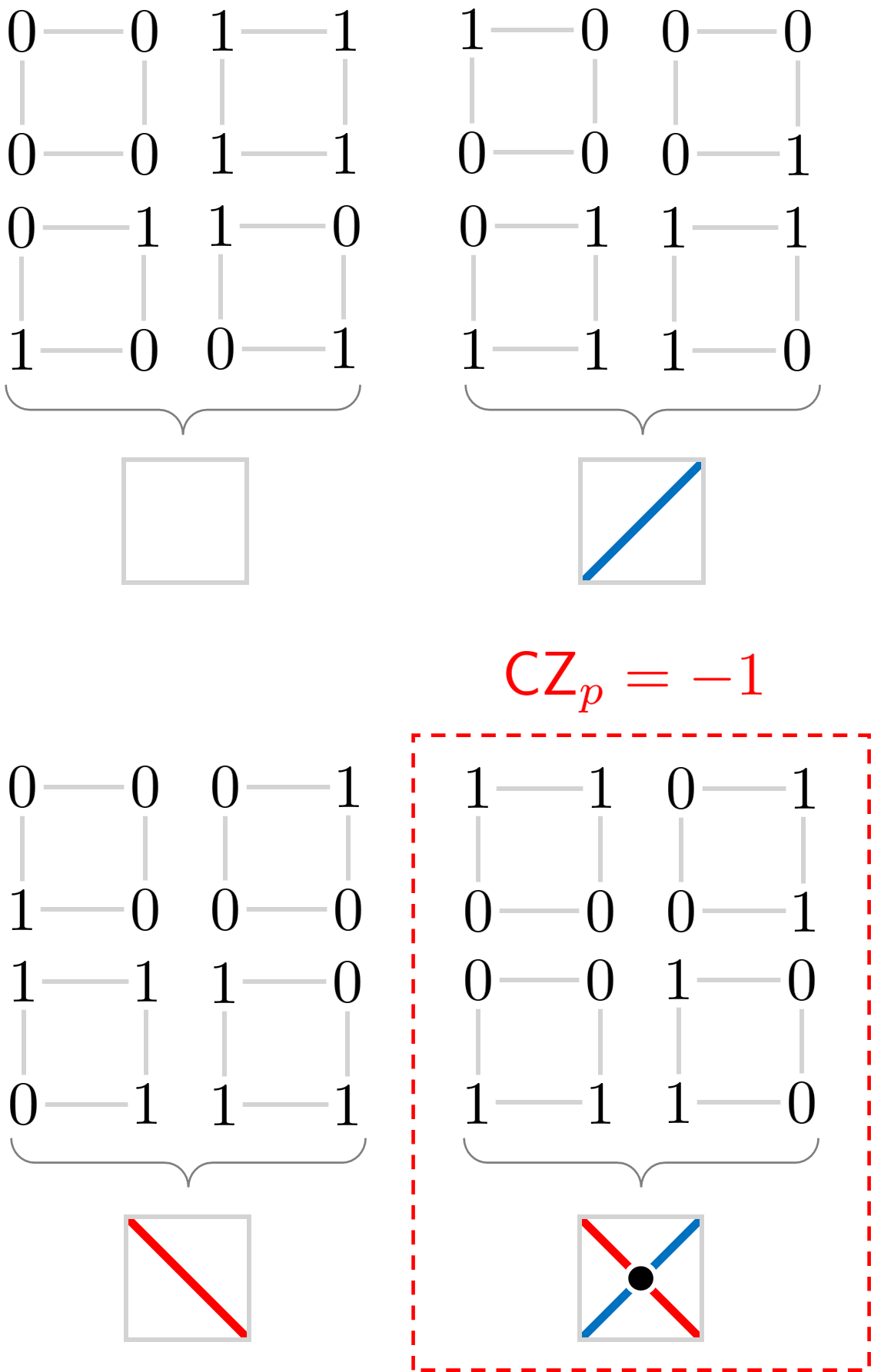}}}
    \hfill
    \subfigure[]{{\label{fig:configurations}\includegraphics[scale=0.295]{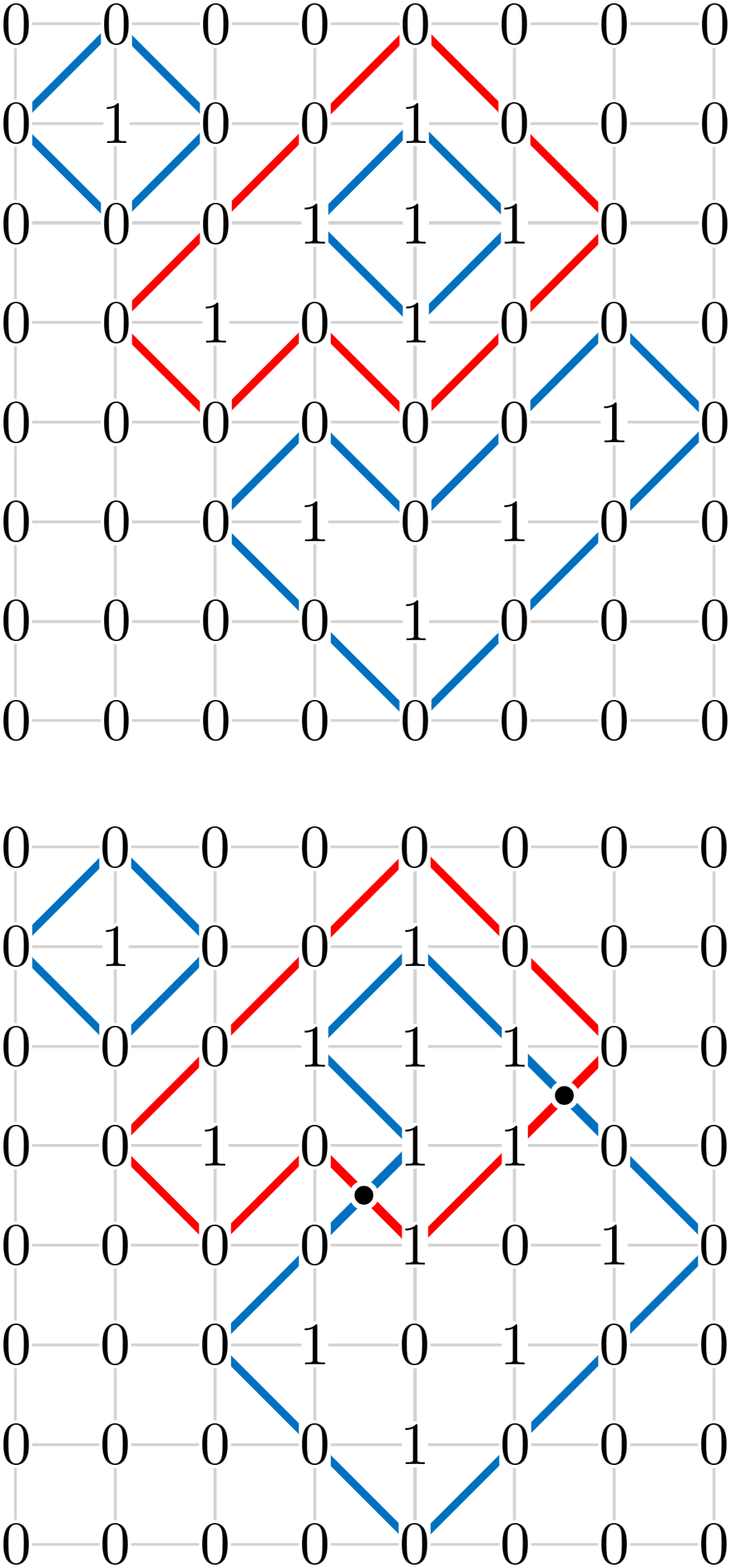}}}
    \hfill
    \subfigure[]{{\label{fig:scars}\includegraphics[scale=0.295]{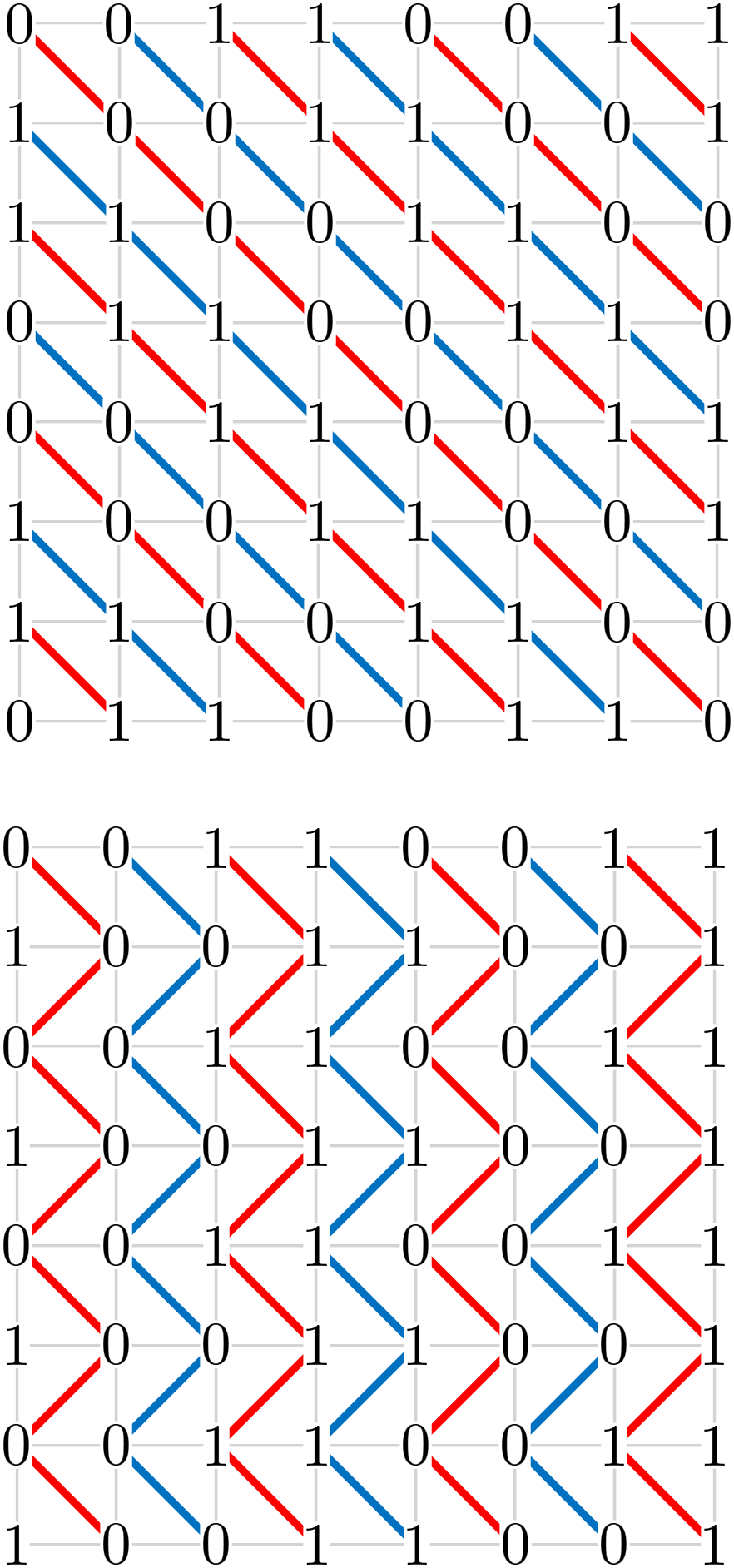}}}
    \subfigure[]{{\label{fig:long_flips}\includegraphics[scale=0.295]{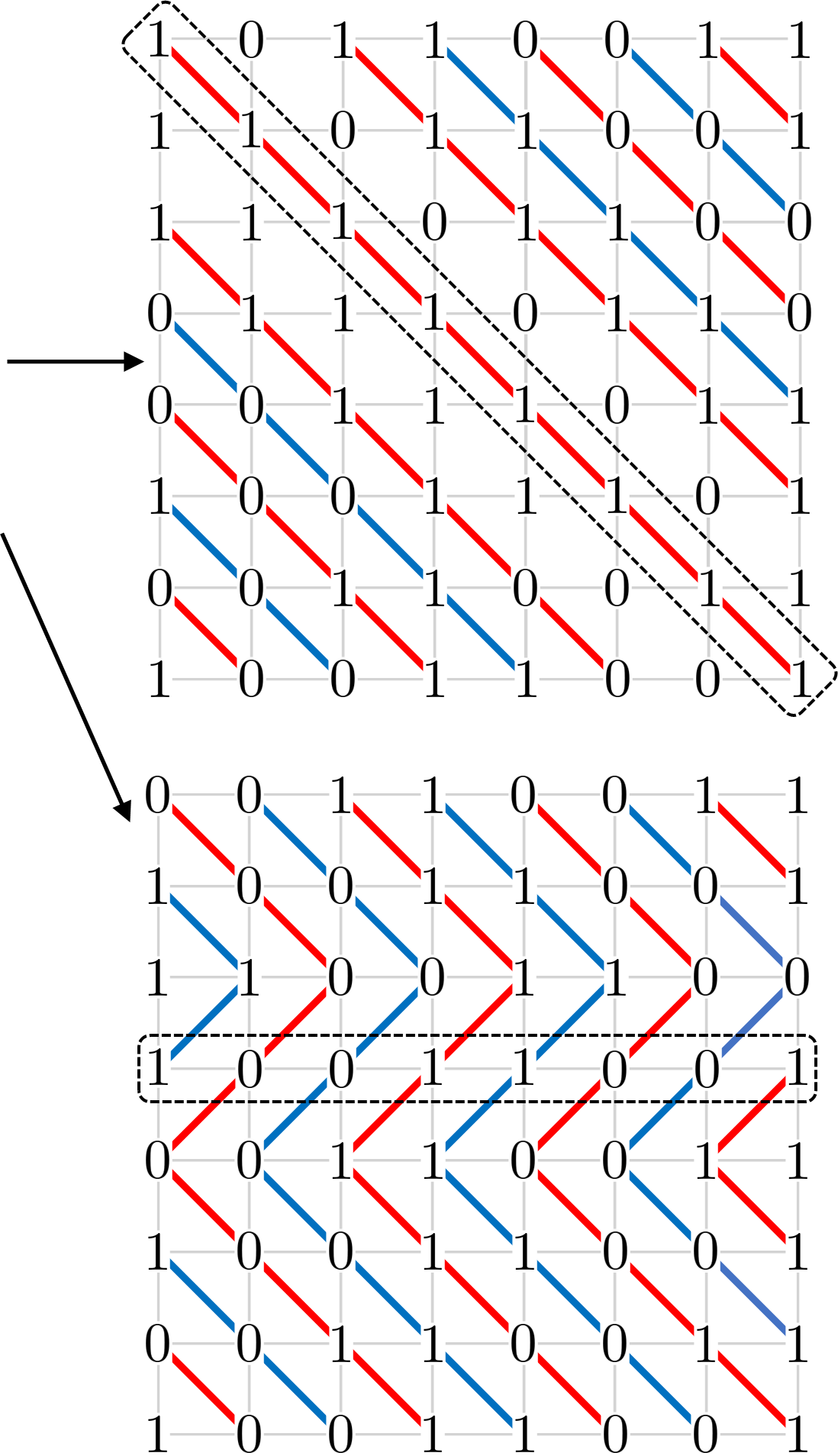}}}
    \caption{(a) The allowed and disallowed configurations of spins around each plaquette. Only the four configurations in the lower right, marked by the dotted box, have an odd number of neighbouring 1's and are therefore given an eigenvalue $\CZ_p=-1$. We group configurations by their image under the duality mapping. The blue (red) lines indicate domain walls of the even (odd) sublattice. (b) Configurations of spins are displayed with their corresponding image under the duality mapping. All configurations in (b), (c), and (d) represent parts of an infinite system. The upper configuration is allowed in the restricted Hilbert space while the lower is not as it contains plaquettes with an odd number of neighbouring 1's, corresponding to the points where loops intersect. (c) Two examples of the scar states, which appear as a foliation of parallel loops in the dual picture. (d) Two examples of states that can be reached by the upper state in (c) by simultaneously flipping all spins in the dashed boxes.}
    \label{fig:duality-and-configs}
\end{figure*}

{\it Model.}---%
We consider a model of spin-$1/2$ particles on an $L\times L$ two-dimensional square lattice. We take periodic boundary conditions in both directions and assume that $L$ is a multiple of 4 \footnote{With open boundary conditions, the scars that we identify can melt from the corners inwards. If $L$ is not divisible by $4$ then the dense packing needed for perfect scars that we describe cannot be attained, although fragmentation may still occur.}. We use the notation $X,Z$ for the Pauli $X$ and $Z$ operators, and denote the basis of $Z$ by the states $|0\rangle, |1\rangle$. The spins interact according to the following Hamiltonian,
\begin{equation} \label{eq:ham}
    H(h) = -J \sum_{ijkl\in\square} \CZ_{ij} \CZ_{jk} \CZ_{kl} \CZ_{il} - h \sum_i X_i.
\end{equation}
Therein, $ijkl\in\square$ represents a set of four spins around a given plaquette (face) of the lattice, ordered clockwise. The operator $\CZ = \id - 2\ket{11}\!\bra{11}$ is the two-body controlled-$Z$ operator, which is diagonal in the $Z$ basis, and gives a minus sign when the two spins are both in the state $|1\rangle$. Let us denote $\CZ_p = \CZ_{ij} \CZ_{jk} \CZ_{kl} \CZ_{il}$ for $i,j,k,l$ being the four sites around plaquette $p$. The $\CZ_p$ interaction can be viewed as a generalized Rydberg interaction. In atomic Rydberg arrays, two neighbouring atoms experience a strong energy shift when they are both in the excited state ($|1\rangle$)~\cite{Browaeys2020rydberg}. This phenomenon, known as the Rydberg blockade, is equivalent to an interaction by the term $\CZ_{ij}$ for each neighbouring pair $i,j$, up to a constant shift. In contrast, $\CZ_p$ gives an energy shift only if there is an odd number of neighbouring sites in the excited state (``neighbouring 1's'') around a given plaquette. Therefore, we have a four-spin parity-dependent interaction which is similar to, but distinct from, the usual two-spin Rydberg interaction. {This interaction can equivalently be expressed in terms of two- and four-body Ising interactions, $\CZ_p=\frac{1}{2}(\id + Z_iZ_k + Z_jZ_l - Z_iZ_jZ_kZ_l$).}

The Hamiltonian in Eq.~\eqref{eq:ham} has a $\Z2 \times \Z2$ symmetry generated by flipping all spins on the even or odd sublattice of the square lattice [a site $i=(x, y)$ is on the even (odd) sublattice if $x+y$ is even (odd)]. Because of this, the interaction $\CZ_p$ depends only on the domain wall variables of the two sublattices. We therefore need only keep track of these domain wall variables [Fig.~\ref{fig:duality}]. The domain walls form two independent sets of closed loops on the two independent sublattices, and the two species of domain wall can intersect with one another on the plaquettes. On each plaquette, there can be no domain wall, a single domain wall between either the odd or the even sites, or a domain wall between both odd and even sites. This gives an effective four-dimensional Hilbert space on each plaquette [Fig.~\ref{fig:duality}]. The interaction $\CZ_p$ acts diagonally on this Hilbert space and gives a factor of $-1$ when there are two domain walls present, i.e., when there is a crossing of loops, and otherwise does nothing. The spin-flip term $\propto \sum_i X_i$ acts in this dual Hilbert space by fluctuating closed-loop configurations locally. This dual picture will be very helpful to visualize the frozen states that we describe in the next section, and to understand their robustness.

{\it Exponentially many perfect scars.}---%
We now consider the limit $J \gg h$. In this limit, the number of domain wall intersections becomes an emergent $\U1$ conserved quantity~\footnote{More precisely, the number of intersections is only conserved up to a quantity of order $\sim h/J$.}, up to a prethermal timescale exponentially \footnote{The exponential scaling is for perturbations more short range than some critical power law \cite{Ho, Nayak1, Nayak2}. For perturbations of longer range, the scaling of the prethermal timescale may be modified. Our arguments should remain valid up to the prethermal timescale, whatever it is.} long in $J/h$~\cite{ADHH}. We work in the
emergent symmetry sector with no intersections, i.e., $\CZ_p=1$ for all plaquettes $p$. This symmetry sector is spanned by product states that have an even number of pairs of neighbouring 1's around every plaquette. Allowed and disallowed configurations are shown in Fig.~\ref{fig:duality}. This symmetry sector is exponentially large in system volume; at the very least, if we put all spins on the even sublattice in the state $|0\rangle$, the state will be in the ground space regardless of the configuration of the odd sublattice, since there will be no neighbouring 1's. In addition to this, configurations with neighbouring 1's are also allowed so long as they respect the parity constraint around each plaquette, see Fig.~\ref{fig:configurations}.
A more quantitative estimate of the sector size is provided by a Pauling estimate~\cite{Pauling1935entropy}, which treats the various local constraints in a mean-field-like manner.
For a given plaquette, 12 of the possible 16 states are permitted by the no-crossing constraint. Once the constraints on adjacent plaquettes are also taken into account (on average), one finds $N_0 \sim 2^{L^2} (12/16)^{N_p} = (3/2)^{L^2}$ states satisfy the constraint, where $N_p = L^2$ is the number of plaquettes. This scaling is in good agreement with exact numerical enumeration of states~\cite{supplement}.

Let us treat $\CZ_p=1$ as a hardcore constraint that defines a restricted Hilbert space. This approximation becomes asymptotically exact in the prethermal limit $h/J \rightarrow 0$. To lowest order in perturbation theory, the effect of the magnetic field projected onto this restricted space is,
\begin{equation}
    H_0 = -h\sum_{i} X_i (P^{0000}_i + P^{1111}_i)
    \, ,
    \label{eq:effective-ham}
\end{equation}
where $P_i^{0000}$ projects the four sites neighbouring $i$ onto the state $\ket{0}$, and similarly for $P_i^{1111}$.
This effective Hamiltonian strongly resembles that of PXP models~\cite{LesanovskyPXP,turner,Turner2018scarred,Michailidis2dPXP,Lin2dPXP}, which arise as effective models in the presence of the Rydberg blockade. PXP models have dynamics given by constrained spin flips, where a spin can only flip if all of its neighbours are in the state $\ket{0}$. Here, we additionally allow the spin flip if all neighbouring spins are in the state $\ket{1}$. This is a consequence of our parity-sensitive interaction in Eq.~\eqref{eq:ham}, since, while the latter spin flip creates new neighbouring 1's (which is not allowed in the conventional Rydberg setup), it conserves the parity of neighbouring 1's around each plaquette. As discussed in Ref.~\cite{Celi2020}, the same lowest-order effective Hamiltonian $H_0$ can be obtained using conventional Rydberg interactions with more than one degree of freedom per lattice site \footnote{Although the lowest-order Hamiltonian is reproduced in Ref.~\cite{Celi2020}, the robustness to higher-orders which we describe shortly will presumably be absent}.

The PXP models are prototypical examples of models that host quantum many-body scars~\cite{serbyn2021review,papic2021review,Moudgalya2022review}. Similarly, the effective Hamiltonian $H_0$~\eqref{eq:effective-ham} also has scars. These scars are simple product states where each site has some neighbours in the state $|0\rangle$ and some in the state $|1\rangle$. Two such states are pictured in Fig.~\ref{fig:scars}. Because no spins are flippable, these states are energy $0$ eigenstates of $H_0$. We remark that $H_0$ is mapped to $-H_0$ by the global application of $Z$, so energy 0 is exactly in the middle of the spectrum of $H_0$ \footnote{We can furthermore show that the ground state of $H_0$ has finite energy density using a variational argument: The state with $|+\rangle = \frac{1}{\sqrt{2}}(|0\rangle + |1\rangle)$ states on even vertices and $|0\rangle$ states on odd vertices has $\langle \psi|\tilde{X}_v|\psi\rangle=1$ for even vertices and $\langle \psi|\tilde{X}_v|\psi\rangle=0$ for odd vertices, giving $\langle \tilde{X}_v\rangle =\frac{1}{2}$ on average, where $\tilde{X}_v = X_v (P^{0000}_v + P^{1111}_v)$ is the conditional spin flip operator. Therefore, the true ground state has energy $E<-h L^2/2$}. Despite this, these states have no entanglement, as they are simply product states, which violates the expectation that states in the middle of the spectrum should have large entanglement. Therefore, we may call them examples of many-body scars. In the dual picture of domain walls, these states look like ``foliations'' of parallel non-contractible loops, as shown in Fig.~\ref{fig:scars}. Because the loops are densely packed, no loop can fluctuate without creating intersections with its neighbouring loops, which would violate our emergent (prethermal) conservation law. 

The states pictured in Fig~\ref{fig:scars} are not the only scar states in this model. In fact, the number of orthogonal scar states grows \emph{exponentially} in linear system size $L$. The other scar states can be constructed in the following way. Observe that the states pictured in Fig.~\ref{fig:scars} consist of the repeated pattern `0011' along every row. In the first row, we can choose to shift this pattern in one of four ways. On each subsequent row, we can independently choose to shift the pattern left or right by one site with respect to the previous row. This generates $2^{L+1}$ states. Rotating the lattice by $90^\circ$ gives $2^{L+1}$ additional states, but they are not all new states. Taking the repeated states into account, there are $2^{L+2}-8$ scar states in total.
The graphical construction makes it clear that these states are all energy 0 eigenstates of $H_0$, and that they are all in the $\CZ_p=1$ sector. In the dual picture, the different scar states come from different ways of putting kinks into the foliated loop pattern.

{\it Absolute robustness of the scar states.}---%
We have shown that, to the lowest order in perturbation theory, we can construct $\sim 2^L$ scar states that have no entanglement and lie in the middle of the energy spectrum. Now, we consider higher orders in perturbation theory. To do this, we need to consider the possibility of a sequence of spin flips that temporarily violates the $\CZ_p=1$ constraint before returning to an allowed state. The dual picture makes it clear that such a process does not exist to \textit{any} finite order in perturbation theory. This is because every fluctuation of loops that is contained within a region with finite radius will inevitably create loop intersections within that region, or its boundary, due to the dense packing of loops. The only process that is allowed is one which pairwise annihilates two loops, or one which puts a kink in all loops across the entire system, see Fig.~\ref{fig:long_flips}. We remark that the former process maps the scar state to a state which can now be acted on by local spin flips [upper state in Fig.~\ref{fig:long_flips}], while the latter maps to another scar state [lower state in Fig.~\ref{fig:long_flips}]. Both processes require simultaneously flipping a number of spins that is proportional to the linear system size $L$, so they only occur at an order of perturbation theory that is also proportional to $L$. Therefore, we say that the scar states pictured in Fig.~\ref{fig:scars} are robust to all finite orders in perturbation theory. We give a more rigorous argument of this robustness in the Supplemental Material (SM)~\cite{supplement}.

Remarkably, these scar states are also robust to arbitrary perturbations of the Hamiltonian. That is to say, in the thermodynamic limit, the scar states will remain eigenstates even in the presence of arbitrary perturbations $V$, in the prethermal limit $\tilde{h}/J \rightarrow 0$, where $\tilde{h}$ is the generalized perturbation strength. We start by noting that, in the prethermal limit, the perturbation must be projected into the symmetry sector with no domain wall intersections. We have already shown that the scar states are an energy 0 eigenstate of $X$ or finite product of $X$ operators, after such projection. On the other hand, since the scar states are product states in the $Z$ basis, they will be eigenstates of any perturbation consisting of $Z$ operators. We note that, unlike the $X$-type perturbations, the $Z$-type perturbations will shift the energy of the scar states away from 0. Since $V$ can always be decomposed into $X$'s, $Z$'s, and their products, we see that the scars are indeed eigenstates of $V$ after projection onto the prethermal symmetry sector, which is valid up to a timescale exponential in $J/h$. We emphasize that we have {\it not} required the perturbations to be symmetry restricted, or to have compact support -- our proof carries through unchanged for perturbations that have long-range tails, and even for perturbations that are fully geometrically nonlocal, as long as they are $k$-body with $k/L\rightarrow 0$ in the thermodynamic limit.

\begin{figure}
    \centering
    \includegraphics[width=0.5\linewidth]{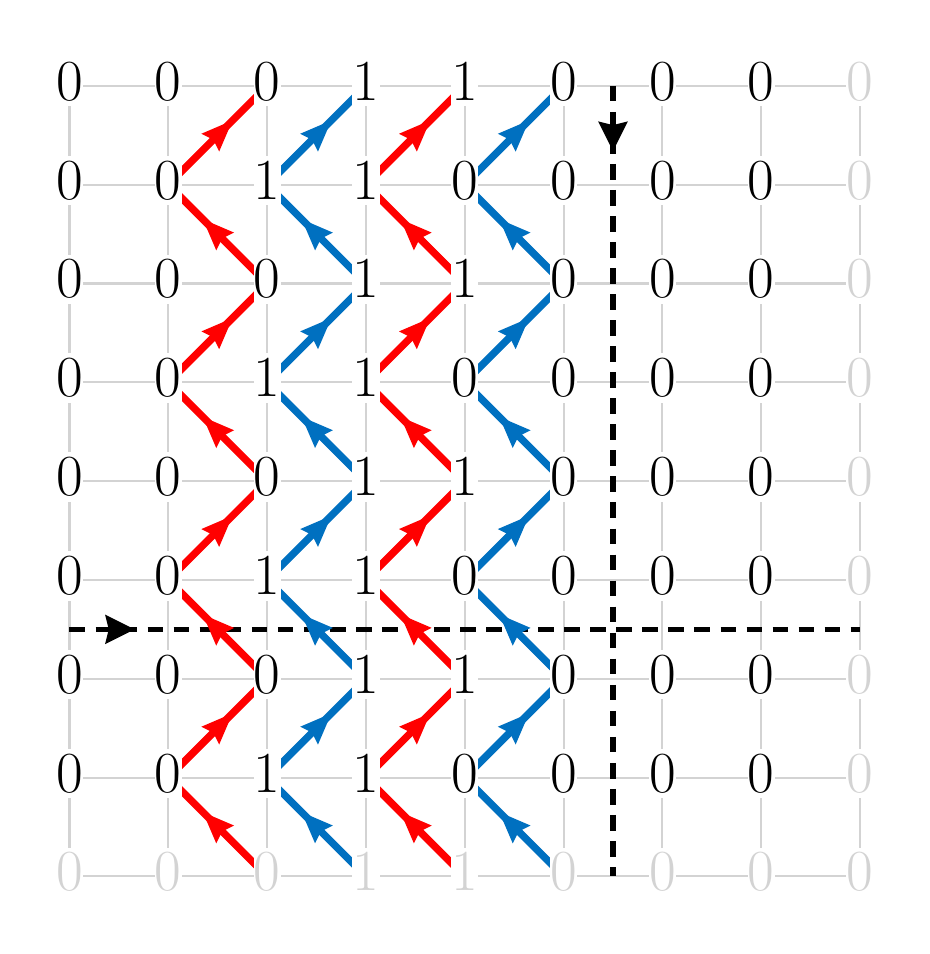}%
    \includegraphics[width=0.5\linewidth]{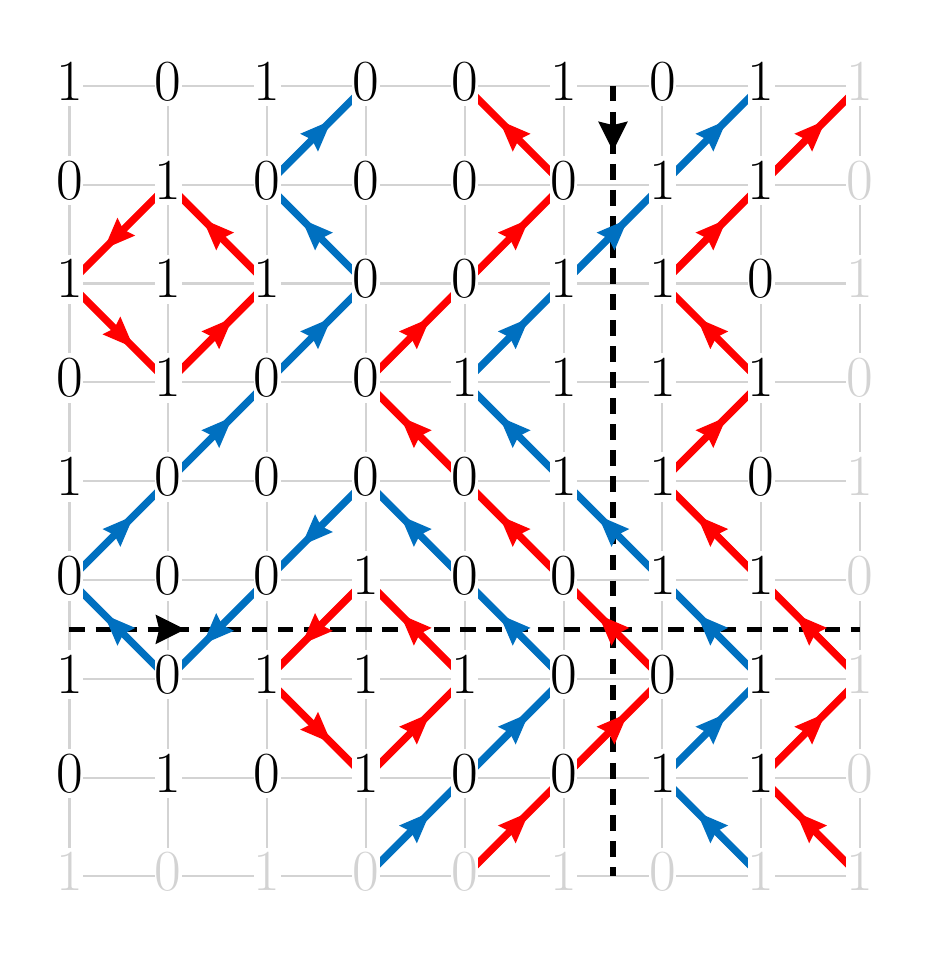}
    \caption{A configuration of spins belonging to one of the `intermediate' sectors. Here, we take an $8\times 8$ lattice with periodic boundary conditions indicated by the grey numbers. 
    The dashed lines indicate rows and columns along which $\mathcal{Q}_y$ and $\mathcal{Q}_x$, respectively, are evaluated. In both cases, we have $\mathcal{Q}_y=4$ and $\mathcal{Q}_x=0$, indicating that these diagrams are in the same sector and are therefore related by a series of local moves.}
    \label{fig:fluctuating-loops}
\end{figure}

{\it Intermediate Krylov sectors.}---%
Now, we investigate the existence of larger isolated (Krylov) sectors of Hilbert space. 
Instead of a dense packing of winding loops [Figs.~\ref{fig:scars}, \ref{fig:long_flips}], consider putting all spins in the state $\ket{0}$ except for three adjacent columns. On these three columns we place the pattern `011' or `110' depending on the parity of the row, as depicted in the left panel of Fig.~\ref{fig:fluctuating-loops}. This produces four adjacent non-contractible domain wall loops of alternating color surrounded by the domain wall vacuum. Under \emph{local} dynamics, these loops can fluctuate, subject to the no-crossing constraint, and contractible loops can be created/destroyed from/into the vacuum, as depicted in the right panel of Fig.~\ref{fig:fluctuating-loops}. Crucially, the no-crossing constraint implies that the number of non-contractible loops of alternating color remains an emergent constant of motion, since two loops of the same color cannot be pairwise annihilated without creating intersections with the loop in between them. More generally, intermediate sectors of Hilbert space can arise from some number of non-contractible loops that wind around any one of the horizontal, vertical, diagonal, or anti-diagonal directions of the torus.

These intermediate sectors of Hilbert space can be understood as symmetry sectors of a $\U1$ 1-form symmetry that emerges due to the $\CZ_p=1$ constraint. To identify this symmetry, we refine the duality mapping from spins to loops to give the loops a well-defined orientation, see SM~\cite{supplement}. Examples of oriented loop patterns are pictured in Fig.~\ref{fig:fluctuating-loops}. For any closed and oriented path $P$ between plaquette centers, we can then compute the net flux $\mathcal{Q}_P$ of loops crossing $P$, where a sign convention can be fixed using the right-hand rule. Any contractible loop will necessarily intersect $P$ twice (in opposite directions) and therefore will not contribute to the net flux. However, if $P$ winds around a non-contractible path on the torus, it is possible for a non-contractible loop to intersect it only once. In particular, if we define $\mathcal{Q}_x$ and $\mathcal{Q}_y$ as the net flux across a non-contractible path along column $x$ and row $y$, as depicted in Fig.~\ref{fig:fluctuating-loops}, then $|\mathcal{Q}_x|$ ($|\mathcal{Q}_y|$) counts the number of loops that wind around the horizontal (vertical) direction of the torus. Importantly, a pair of adjacent non-contractible loops of the same color turn out to have opposite orientations. Therefore, only the loops of alternating color, which cannot be annihilated by local dynamics without introducing crossings, will contribute to $\mathcal{Q}_{x/y}$. If a domain wall loop winds diagonally around the torus, it will contribute to both $\mathcal{Q}_x$ and $\mathcal{Q}_y$. Since $\mathcal{Q}_P=0$ for any contractible path $P$, the paths along which $\mathcal{Q}_{x/y}$ are evaluated can be arbitrarily locally deformed without changing their value, hence they should be considered 1-form symmetry charges.

Different signs of $\mathcal{Q}_{x/y}$ also label disconnected sectors of Hilbert space. In the trivial sector, \textit{i.e.} the sector with $\mathcal{Q}_{x}=\mathcal{Q}_{y}=0$, we can flip all spins on a given sublattice by nucleating a small domain wall on that sublattice and growing it to cover the entire system. For any other sector, this is not possible, since the domain wall will not be able to pass through the non-contractible loops corresponding to the other sublattice. We can, however, flip every spin by sweeping a pair of non-contractible loops across the entire system. Indeed, flipping every other spin turns out to reverse the orientation of all loops, sending $\mathcal{Q}_{x/y}\rightarrow -\mathcal{Q}_{x/y}$. When both $\mathcal{Q}_x$ and $\mathcal{Q}_y$ are non-zero, their relative sign indicates whether the non-contractible domain walls wrap the torus around the diagonal or anti-diagonal. 

Geometrically, it is clear that local fluctuations in the domain walls cannot change $\mathcal{Q}_{x/y}$, so they are constants of motion. Small-scale numerical studies of the connectivity of Hilbert space under the dynamics of $H_0$ support the claim that the values of $\mathcal{Q}_{x/y}$ uniquely label \emph{all} intermediate sectors of Hilbert space, where domain walls are not fully packed \cite{supplement}. These sectors can thus be viewed as symmetry sectors of the one-form symmetry. However the scar states cannot be viewed in this way, as there are indeed only $\sim L$ different values of $\mathcal{Q}_{x/y}$, whereas the number of scar states is $\sim \exp(L)$. Therefore, the $\U1$ 1-form symmetry sectors are {\it not} sufficient to uniquely label all the fragments of Hilbert space. Finally, we note that this $\U1$ 1-form symmetry is indeed an \textit{emergent} symmetry in the $\CZ_p=1$ sector, as loop crossings turn out to act as sources or sinks of oriented flux \cite{supplement}.

{\it Hydrodynamics}---%
We now discuss dynamics from simple non-scar initial states. Within the $\CZ_p=1$ sector we show (see SM~\cite{supplement}) that this is well described by magnetohydrodynamics of the emergent one-form $\U1$ symmetry \cite{fractonMHD}. This analytic expectation may be confirmed using the automaton Monte Carlo technique~\cite{IVN} and single spin flip dynamics. Meanwhile, in sectors with intersections (a nonzero number of $\CZ_p = -1$) we find using the automaton Monte Carlo technique (see SM~\cite{supplement}) that starting from an `infinite temperature' initial condition (i.e., a typical state of the spins, containing many intersections), the long-time limit of the subsequent dynamics is characterized by isotropic diffusion of the intersections, which (we recall) are conserved up to the prethermal timescale.

{\it Conclusions}.---%
We have shown how one may obtain ergodicity breaking via Hilbert space fragmentation that is provably robust to arbitrary perturbations, as long as the perturbations are $k$-local with $k/L \rightarrow 0$ in the thermodynamic limit, where $L$ is linear system size \footnote{We note in passing that our results are presented strictly in the Schrieffer-Wolff transformed basis, and in the `bare' basis there can be non-trivial dynamics, see \cite{supplement}}. Our approach relies on the existence of a single emergent (prethermal) $\U1$ conservation law. We have demonstrated the existence of a number of frozen configurations (perfect scars) exponential in $L$, as well as larger subsectors labelled by an emergent topological winding number. The results arise from a `generalized Rydberg constraint', whose lowest-order dynamics are accessible in the near term in synthetic quantum systems~\cite{Celi2020}. 

Our results open a new chapter for the study of ergodicity breaking. For instance, one may wonder whether the degree of ergodicity breaking here is sufficient to define local integrals of motion~\cite{GeraedtsLIOM, SLIOM}. More generally, our work may provide a new avenue for the design of ergodicity breaking models, which may function as robust memories, and may also be accessible in near-term quantum simulators. Most broadly, they may prompt a re-evaluation of the necessary desiderata for ergodicity and quantum thermalization.

\let\oldaddcontentsline\addcontentsline
\renewcommand{\addcontentsline}[3]{}

\acknowledgments{RN and OH are supported by the Air Force Office of Scientific Research under award number FA9550-20-1-0222. RN also thanks the Simons Foundation for a Simons Fellowship in Theoretical Physics. DTS is supported by the Simons Collaboration on Ultra-Quantum Matter, which is a grant from the Simons Foundation (651440).}

\bibliography{scars_bib.bib}

\begin{thebibliography}{51}%
\makeatletter
\providecommand \@ifxundefined [1]{%
 \@ifx{#1\undefined}
}%
\providecommand \@ifnum [1]{%
 \ifnum #1\expandafter \@firstoftwo
 \else \expandafter \@secondoftwo
 \fi
}%
\providecommand \@ifx [1]{%
 \ifx #1\expandafter \@firstoftwo
 \else \expandafter \@secondoftwo
 \fi
}%
\providecommand \natexlab [1]{#1}%
\providecommand \enquote  [1]{``#1''}%
\providecommand \bibnamefont  [1]{#1}%
\providecommand \bibfnamefont [1]{#1}%
\providecommand \citenamefont [1]{#1}%
\providecommand \href@noop [0]{\@secondoftwo}%
\providecommand \href [0]{\begingroup \@sanitize@url \@href}%
\providecommand \@href[1]{\@@startlink{#1}\@@href}%
\providecommand \@@href[1]{\endgroup#1\@@endlink}%
\providecommand \@sanitize@url [0]{\catcode `\\12\catcode `\$12\catcode
  `\&12\catcode `\#12\catcode `\^12\catcode `\_12\catcode `\%12\relax}%
\providecommand \@@startlink[1]{}%
\providecommand \@@endlink[0]{}%
\providecommand \url  [0]{\begingroup\@sanitize@url \@url }%
\providecommand \@url [1]{\endgroup\@href {#1}{\urlprefix }}%
\providecommand \urlprefix  [0]{URL }%
\providecommand \Eprint [0]{\href }%
\providecommand \doibase [0]{https://doi.org/}%
\providecommand \selectlanguage [0]{\@gobble}%
\providecommand \bibinfo  [0]{\@secondoftwo}%
\providecommand \bibfield  [0]{\@secondoftwo}%
\providecommand \translation [1]{[#1]}%
\providecommand \BibitemOpen [0]{}%
\providecommand \bibitemStop [0]{}%
\providecommand \bibitemNoStop [0]{.\EOS\space}%
\providecommand \EOS [0]{\spacefactor3000\relax}%
\providecommand \BibitemShut  [1]{\csname bibitem#1\endcsname}%
\let\auto@bib@innerbib\@empty
\bibitem [{\citenamefont {Baxter}(2016)}]{Baxter}%
  \BibitemOpen
  \bibfield  {author} {\bibinfo {author} {\bibfnamefont {R.~J.}\ \bibnamefont
  {Baxter}},\ }\href@noop {} {\emph {\bibinfo {title} {Exactly solved models in
  statistical mechanics}}}\ (\bibinfo  {publisher} {Elsevier},\ \bibinfo {year}
  {2016})\BibitemShut {NoStop}%
\bibitem [{\citenamefont {Nandkishore}\ and\ \citenamefont
  {Huse}(2015)}]{mblarcmp}%
  \BibitemOpen
  \bibfield  {author} {\bibinfo {author} {\bibfnamefont {R.}~\bibnamefont
  {Nandkishore}}\ and\ \bibinfo {author} {\bibfnamefont {D.~A.}\ \bibnamefont
  {Huse}},\ }\bibfield  {title} {\bibinfo {title} {Many-body localization and
  thermalization in quantum statistical mechanics},\ }\href
  {https://doi.org/10.1146/annurev-conmatphys-031214-014726} {\bibfield
  {journal} {\bibinfo  {journal} {Annu. Rev. Condens. Matter Phys.}\ }\textbf
  {\bibinfo {volume} {6}},\ \bibinfo {pages} {15} (\bibinfo {year}
  {2015})}\BibitemShut {NoStop}%
\bibitem [{\citenamefont {Abanin}\ \emph {et~al.}(2019)\citenamefont {Abanin},
  \citenamefont {Altman}, \citenamefont {Bloch},\ and\ \citenamefont
  {Serbyn}}]{mblrmp}%
  \BibitemOpen
  \bibfield  {author} {\bibinfo {author} {\bibfnamefont {D.~A.}\ \bibnamefont
  {Abanin}}, \bibinfo {author} {\bibfnamefont {E.}~\bibnamefont {Altman}},
  \bibinfo {author} {\bibfnamefont {I.}~\bibnamefont {Bloch}},\ and\ \bibinfo
  {author} {\bibfnamefont {M.}~\bibnamefont {Serbyn}},\ }\bibfield  {title}
  {\bibinfo {title} {Colloquium: Many-body localization, thermalization, and
  entanglement},\ }\href {https://doi.org/10.1103/RevModPhys.91.021001}
  {\bibfield  {journal} {\bibinfo  {journal} {Reviews of Modern Physics}\
  }\textbf {\bibinfo {volume} {91}},\ \bibinfo {pages} {021001} (\bibinfo
  {year} {2019})}\BibitemShut {NoStop}%
\bibitem [{\citenamefont {Shiraishi}\ and\ \citenamefont
  {Mori}(2017)}]{shiraishimori}%
  \BibitemOpen
  \bibfield  {author} {\bibinfo {author} {\bibfnamefont {N.}~\bibnamefont
  {Shiraishi}}\ and\ \bibinfo {author} {\bibfnamefont {T.}~\bibnamefont
  {Mori}},\ }\bibfield  {title} {\bibinfo {title} {Systematic construction of
  counterexamples to the eigenstate thermalization hypothesis},\ }\href
  {https://doi.org/10.1103/PhysRevLett.119.030601} {\bibfield  {journal}
  {\bibinfo  {journal} {Phys. Rev. Lett.}\ }\textbf {\bibinfo {volume} {119}},\
  \bibinfo {pages} {030601} (\bibinfo {year} {2017})}\BibitemShut {NoStop}%
\bibitem [{\citenamefont {Moudgalya}\ \emph {et~al.}(2018)\citenamefont
  {Moudgalya}, \citenamefont {Regnault},\ and\ \citenamefont
  {Bernevig}}]{aklt}%
  \BibitemOpen
  \bibfield  {author} {\bibinfo {author} {\bibfnamefont {S.}~\bibnamefont
  {Moudgalya}}, \bibinfo {author} {\bibfnamefont {N.}~\bibnamefont
  {Regnault}},\ and\ \bibinfo {author} {\bibfnamefont {B.~A.}\ \bibnamefont
  {Bernevig}},\ }\bibfield  {title} {\bibinfo {title} {Entanglement of exact
  excited states of affleck-kennedy-lieb-tasaki models: Exact results,
  many-body scars, and violation of the strong eigenstate thermalization
  hypothesis},\ }\href {https://doi.org/10.1103/PhysRevB.98.235156} {\bibfield
  {journal} {\bibinfo  {journal} {Physical Review B}\ }\textbf {\bibinfo
  {volume} {98}},\ \bibinfo {pages} {235156} (\bibinfo {year}
  {2018})}\BibitemShut {NoStop}%
\bibitem [{\citenamefont {Turner}\ \emph
  {et~al.}(2018{\natexlab{a}})\citenamefont {Turner}, \citenamefont
  {Michailidis}, \citenamefont {Abanin}, \citenamefont {Serbyn},\ and\
  \citenamefont {Papi{\'c}}}]{turner}%
  \BibitemOpen
  \bibfield  {author} {\bibinfo {author} {\bibfnamefont {C.~J.}\ \bibnamefont
  {Turner}}, \bibinfo {author} {\bibfnamefont {A.~A.}\ \bibnamefont
  {Michailidis}}, \bibinfo {author} {\bibfnamefont {D.~A.}\ \bibnamefont
  {Abanin}}, \bibinfo {author} {\bibfnamefont {M.}~\bibnamefont {Serbyn}},\
  and\ \bibinfo {author} {\bibfnamefont {Z.}~\bibnamefont {Papi{\'c}}},\
  }\bibfield  {title} {\bibinfo {title} {Weak ergodicity breaking from quantum
  many-body scars},\ }\href {https://doi.org/10.1038/s41567-018-0137-5}
  {\bibfield  {journal} {\bibinfo  {journal} {Nature Physics}\ }\textbf
  {\bibinfo {volume} {14}},\ \bibinfo {pages} {745} (\bibinfo {year}
  {2018}{\natexlab{a}})}\BibitemShut {NoStop}%
\bibitem [{\citenamefont {Chandran}\ \emph {et~al.}(2022)\citenamefont
  {Chandran}, \citenamefont {Iadecola}, \citenamefont {Khemani},\ and\
  \citenamefont {Moessner}}]{scarsarcmp}%
  \BibitemOpen
  \bibfield  {author} {\bibinfo {author} {\bibfnamefont {A.}~\bibnamefont
  {Chandran}}, \bibinfo {author} {\bibfnamefont {T.}~\bibnamefont {Iadecola}},
  \bibinfo {author} {\bibfnamefont {V.}~\bibnamefont {Khemani}},\ and\ \bibinfo
  {author} {\bibfnamefont {R.}~\bibnamefont {Moessner}},\ }\bibfield  {title}
  {\bibinfo {title} {Quantum many-body scars: A quasiparticle perspective},\
  }\bibfield  {journal} {\bibinfo  {journal} {arXiv preprint arXiv:2206.11528}\
  }\href {https://doi.org/10.48550/arXiv.2206.11528}
  {10.48550/arXiv.2206.11528} (\bibinfo {year} {2022})\BibitemShut {NoStop}%
\bibitem [{\citenamefont {Pai}\ \emph {et~al.}(2019)\citenamefont {Pai},
  \citenamefont {Pretko},\ and\ \citenamefont {Nandkishore}}]{ppn}%
  \BibitemOpen
  \bibfield  {author} {\bibinfo {author} {\bibfnamefont {S.}~\bibnamefont
  {Pai}}, \bibinfo {author} {\bibfnamefont {M.}~\bibnamefont {Pretko}},\ and\
  \bibinfo {author} {\bibfnamefont {R.~M.}\ \bibnamefont {Nandkishore}},\
  }\bibfield  {title} {\bibinfo {title} {Localization in fractonic random
  circuits},\ }\href {https://doi.org/10.1103/PhysRevX.9.021003} {\bibfield
  {journal} {\bibinfo  {journal} {Physical Review X}\ }\textbf {\bibinfo
  {volume} {9}},\ \bibinfo {pages} {021003} (\bibinfo {year}
  {2019})}\BibitemShut {NoStop}%
\bibitem [{\citenamefont {Khemani}\ \emph {et~al.}(2020)\citenamefont
  {Khemani}, \citenamefont {Hermele},\ and\ \citenamefont {Nandkishore}}]{KHN}%
  \BibitemOpen
  \bibfield  {author} {\bibinfo {author} {\bibfnamefont {V.}~\bibnamefont
  {Khemani}}, \bibinfo {author} {\bibfnamefont {M.}~\bibnamefont {Hermele}},\
  and\ \bibinfo {author} {\bibfnamefont {R.}~\bibnamefont {Nandkishore}},\
  }\bibfield  {title} {\bibinfo {title} {Localization from hilbert space
  shattering: From theory to physical realizations},\ }\href
  {https://doi.org/10.1103/PhysRevB.101.174204} {\bibfield  {journal} {\bibinfo
   {journal} {Physical Review B}\ }\textbf {\bibinfo {volume} {101}},\ \bibinfo
  {pages} {174204} (\bibinfo {year} {2020})}\BibitemShut {NoStop}%
\bibitem [{\citenamefont {Sala}\ \emph {et~al.}(2020)\citenamefont {Sala},
  \citenamefont {Rakovszky}, \citenamefont {Verresen}, \citenamefont {Knap},\
  and\ \citenamefont {Pollmann}}]{Sala}%
  \BibitemOpen
  \bibfield  {author} {\bibinfo {author} {\bibfnamefont {P.}~\bibnamefont
  {Sala}}, \bibinfo {author} {\bibfnamefont {T.}~\bibnamefont {Rakovszky}},
  \bibinfo {author} {\bibfnamefont {R.}~\bibnamefont {Verresen}}, \bibinfo
  {author} {\bibfnamefont {M.}~\bibnamefont {Knap}},\ and\ \bibinfo {author}
  {\bibfnamefont {F.}~\bibnamefont {Pollmann}},\ }\bibfield  {title} {\bibinfo
  {title} {Ergodicity breaking arising from hilbert space fragmentation in
  dipole-conserving hamiltonians},\ }\href
  {https://doi.org/10.1103/PhysRevX.10.011047} {\bibfield  {journal} {\bibinfo
  {journal} {Physical Review X}\ }\textbf {\bibinfo {volume} {10}},\ \bibinfo
  {pages} {011047} (\bibinfo {year} {2020})}\BibitemShut {NoStop}%
\bibitem [{\citenamefont {Moudgalya}\ \emph
  {et~al.}(2022{\natexlab{a}})\citenamefont {Moudgalya}, \citenamefont {Prem},
  \citenamefont {Nandkishore}, \citenamefont {Regnault},\ and\ \citenamefont
  {Bernevig}}]{Moudgalya}%
  \BibitemOpen
  \bibfield  {author} {\bibinfo {author} {\bibfnamefont {S.}~\bibnamefont
  {Moudgalya}}, \bibinfo {author} {\bibfnamefont {A.}~\bibnamefont {Prem}},
  \bibinfo {author} {\bibfnamefont {R.}~\bibnamefont {Nandkishore}}, \bibinfo
  {author} {\bibfnamefont {N.}~\bibnamefont {Regnault}},\ and\ \bibinfo
  {author} {\bibfnamefont {B.~A.}\ \bibnamefont {Bernevig}},\ }\bibfield
  {title} {\bibinfo {title} {Thermalization and its absence within krylov
  subspaces of a constrained hamiltonian},\ }in\ \href
  {https://doi.org/10.1142/9789811231711_0009} {\emph {\bibinfo {booktitle}
  {Memorial Volume for Shoucheng Zhang}}}\ (\bibinfo  {publisher} {World
  Scientific},\ \bibinfo {year} {2022})\ pp.\ \bibinfo {pages}
  {147--209}\BibitemShut {NoStop}%
\bibitem [{\citenamefont {Rakovszky}\ \emph {et~al.}(2020)\citenamefont
  {Rakovszky}, \citenamefont {Sala}, \citenamefont {Verresen}, \citenamefont
  {Knap},\ and\ \citenamefont {Pollmann}}]{SLIOM}%
  \BibitemOpen
  \bibfield  {author} {\bibinfo {author} {\bibfnamefont {T.}~\bibnamefont
  {Rakovszky}}, \bibinfo {author} {\bibfnamefont {P.}~\bibnamefont {Sala}},
  \bibinfo {author} {\bibfnamefont {R.}~\bibnamefont {Verresen}}, \bibinfo
  {author} {\bibfnamefont {M.}~\bibnamefont {Knap}},\ and\ \bibinfo {author}
  {\bibfnamefont {F.}~\bibnamefont {Pollmann}},\ }\bibfield  {title} {\bibinfo
  {title} {Statistical localization: From strong fragmentation to strong edge
  modes},\ }\href {https://doi.org/10.1103/PhysRevB.101.125126} {\bibfield
  {journal} {\bibinfo  {journal} {Physical Review B}\ }\textbf {\bibinfo
  {volume} {101}},\ \bibinfo {pages} {125126} (\bibinfo {year}
  {2020})}\BibitemShut {NoStop}%
\bibitem [{\citenamefont {Yang}\ \emph {et~al.}(2020)\citenamefont {Yang},
  \citenamefont {Liu}, \citenamefont {Gorshkov},\ and\ \citenamefont
  {Iadecola}}]{Yang2020confinement}%
  \BibitemOpen
  \bibfield  {author} {\bibinfo {author} {\bibfnamefont {Z.-C.}\ \bibnamefont
  {Yang}}, \bibinfo {author} {\bibfnamefont {F.}~\bibnamefont {Liu}}, \bibinfo
  {author} {\bibfnamefont {A.~V.}\ \bibnamefont {Gorshkov}},\ and\ \bibinfo
  {author} {\bibfnamefont {T.}~\bibnamefont {Iadecola}},\ }\bibfield  {title}
  {\bibinfo {title} {Hilbert-space fragmentation from strict confinement},\
  }\href {https://doi.org/10.1103/PhysRevLett.124.207602} {\bibfield  {journal}
  {\bibinfo  {journal} {Phys. Rev. Lett.}\ }\textbf {\bibinfo {volume} {124}},\
  \bibinfo {pages} {207602} (\bibinfo {year} {2020})}\BibitemShut {NoStop}%
\bibitem [{\citenamefont {Khudorozhkov}\ \emph {et~al.}(2021)\citenamefont
  {Khudorozhkov}, \citenamefont {Tiwari}, \citenamefont {Chamon},\ and\
  \citenamefont {Neupert}}]{chamon}%
  \BibitemOpen
  \bibfield  {author} {\bibinfo {author} {\bibfnamefont {A.}~\bibnamefont
  {Khudorozhkov}}, \bibinfo {author} {\bibfnamefont {A.}~\bibnamefont
  {Tiwari}}, \bibinfo {author} {\bibfnamefont {C.}~\bibnamefont {Chamon}},\
  and\ \bibinfo {author} {\bibfnamefont {T.}~\bibnamefont {Neupert}},\
  }\bibfield  {title} {\bibinfo {title} {Hilbert space fragmentation in a 2d
  quantum spin system with subsystem symmetries},\ }\bibfield  {journal}
  {\bibinfo  {journal} {arXiv preprint arXiv:2107.09690}\ }\href
  {https://doi.org/10.48550/arXiv.2107.09690} {10.48550/arXiv.2107.09690}
  (\bibinfo {year} {2021})\BibitemShut {NoStop}%
\bibitem [{\citenamefont {Moudgalya}\ and\ \citenamefont
  {Motrunich}(2022)}]{commutant}%
  \BibitemOpen
  \bibfield  {author} {\bibinfo {author} {\bibfnamefont {S.}~\bibnamefont
  {Moudgalya}}\ and\ \bibinfo {author} {\bibfnamefont {O.~I.}\ \bibnamefont
  {Motrunich}},\ }\bibfield  {title} {\bibinfo {title} {Hilbert space
  fragmentation and commutant algebras},\ }\href
  {https://doi.org/10.1103/PhysRevX.12.011050} {\bibfield  {journal} {\bibinfo
  {journal} {Physical Review X}\ }\textbf {\bibinfo {volume} {12}},\ \bibinfo
  {pages} {011050} (\bibinfo {year} {2022})}\BibitemShut {NoStop}%
\bibitem [{\citenamefont {Yoshinaga}\ \emph {et~al.}(2022)\citenamefont
  {Yoshinaga}, \citenamefont {Hakoshima}, \citenamefont {Imoto}, \citenamefont
  {Matsuzaki},\ and\ \citenamefont {Hamazaki}}]{YoshinagaIsing}%
  \BibitemOpen
  \bibfield  {author} {\bibinfo {author} {\bibfnamefont {A.}~\bibnamefont
  {Yoshinaga}}, \bibinfo {author} {\bibfnamefont {H.}~\bibnamefont
  {Hakoshima}}, \bibinfo {author} {\bibfnamefont {T.}~\bibnamefont {Imoto}},
  \bibinfo {author} {\bibfnamefont {Y.}~\bibnamefont {Matsuzaki}},\ and\
  \bibinfo {author} {\bibfnamefont {R.}~\bibnamefont {Hamazaki}},\ }\bibfield
  {title} {\bibinfo {title} {Emergence of hilbert space fragmentation in ising
  models with a weak transverse field},\ }\href
  {https://doi.org/10.1103/PhysRevLett.129.090602} {\bibfield  {journal}
  {\bibinfo  {journal} {Phys. Rev. Lett.}\ }\textbf {\bibinfo {volume} {129}},\
  \bibinfo {pages} {090602} (\bibinfo {year} {2022})}\BibitemShut {NoStop}%
\bibitem [{\citenamefont {Hart}\ and\ \citenamefont {Nandkishore}(2022)}]{hn}%
  \BibitemOpen
  \bibfield  {author} {\bibinfo {author} {\bibfnamefont {O.}~\bibnamefont
  {Hart}}\ and\ \bibinfo {author} {\bibfnamefont {R.}~\bibnamefont
  {Nandkishore}},\ }\bibfield  {title} {\bibinfo {title} {Hilbert space
  shattering and dynamical freezing in the quantum ising model},\ }\bibfield
  {journal} {\bibinfo  {journal} {arXiv preprint arXiv:2203.06188}\ }\href
  {https://doi.org/10.48550/arXiv.2203.06188} {10.48550/arXiv.2203.06188}
  (\bibinfo {year} {2022})\BibitemShut {NoStop}%
\bibitem [{\citenamefont {Imbrie}(2016)}]{imbrie}%
  \BibitemOpen
  \bibfield  {author} {\bibinfo {author} {\bibfnamefont {J.~Z.}\ \bibnamefont
  {Imbrie}},\ }\bibfield  {title} {\bibinfo {title} {On many-body localization
  for quantum spin chains},\ }\href {https://doi.org/10.1007/s10955-016-1508-x}
  {\bibfield  {journal} {\bibinfo  {journal} {Journal of Statistical Physics}\
  }\textbf {\bibinfo {volume} {163}},\ \bibinfo {pages} {998} (\bibinfo {year}
  {2016})}\BibitemShut {NoStop}%
\bibitem [{\citenamefont {Sels}\ and\ \citenamefont
  {Polkovnikov}(2021)}]{selspolkovnikov}%
  \BibitemOpen
  \bibfield  {author} {\bibinfo {author} {\bibfnamefont {D.}~\bibnamefont
  {Sels}}\ and\ \bibinfo {author} {\bibfnamefont {A.}~\bibnamefont
  {Polkovnikov}},\ }\bibfield  {title} {\bibinfo {title} {Dynamical obstruction
  to localization in a disordered spin chain},\ }\href
  {https://doi.org/10.1103/PhysRevE.104.054105} {\bibfield  {journal} {\bibinfo
   {journal} {Physical Review E}\ }\textbf {\bibinfo {volume} {104}},\ \bibinfo
  {pages} {054105} (\bibinfo {year} {2021})}\BibitemShut {NoStop}%
\bibitem [{\citenamefont {Abanin}\ \emph {et~al.}(2017)\citenamefont {Abanin},
  \citenamefont {De~Roeck}, \citenamefont {Ho},\ and\ \citenamefont
  {Huveneers}}]{ADHH}%
  \BibitemOpen
  \bibfield  {author} {\bibinfo {author} {\bibfnamefont {D.}~\bibnamefont
  {Abanin}}, \bibinfo {author} {\bibfnamefont {W.}~\bibnamefont {De~Roeck}},
  \bibinfo {author} {\bibfnamefont {W.~W.}\ \bibnamefont {Ho}},\ and\ \bibinfo
  {author} {\bibfnamefont {F.}~\bibnamefont {Huveneers}},\ }\bibfield  {title}
  {\bibinfo {title} {A rigorous theory of many-body prethermalization for
  periodically driven and closed quantum systems},\ }\href
  {https://doi.org/10.1007/s00220-017-2930-x} {\bibfield  {journal} {\bibinfo
  {journal} {Communications in Mathematical Physics}\ }\textbf {\bibinfo
  {volume} {354}},\ \bibinfo {pages} {809} (\bibinfo {year}
  {2017})}\BibitemShut {NoStop}%
\bibitem [{\citenamefont {von Keyserlingk}\ \emph {et~al.}(2016)\citenamefont
  {von Keyserlingk}, \citenamefont {Khemani},\ and\ \citenamefont
  {Sondhi}}]{Keyserlingk}%
  \BibitemOpen
  \bibfield  {author} {\bibinfo {author} {\bibfnamefont {C.~W.}\ \bibnamefont
  {von Keyserlingk}}, \bibinfo {author} {\bibfnamefont {V.}~\bibnamefont
  {Khemani}},\ and\ \bibinfo {author} {\bibfnamefont {S.~L.}\ \bibnamefont
  {Sondhi}},\ }\bibfield  {title} {\bibinfo {title} {Absolute stability and
  spatiotemporal long-range order in floquet systems},\ }\href
  {https://doi.org/10.1103/PhysRevB.94.085112} {\bibfield  {journal} {\bibinfo
  {journal} {Phys. Rev. B}\ }\textbf {\bibinfo {volume} {94}},\ \bibinfo
  {pages} {085112} (\bibinfo {year} {2016})}\BibitemShut {NoStop}%
\bibitem [{Note1()}]{Note1}%
  \BibitemOpen
  \bibinfo {note} {With open boundary conditions, the scars that we identify
  can melt from the corners inwards. If $L$ is not divisible by $4$ then the
  dense packing needed for perfect scars that we describe cannot be attained,
  although fragmentation may still occur.}\BibitemShut {Stop}%
\bibitem [{\citenamefont {Browaeys}\ and\ \citenamefont
  {Lahaye}(2020)}]{Browaeys2020rydberg}%
  \BibitemOpen
  \bibfield  {author} {\bibinfo {author} {\bibfnamefont {A.}~\bibnamefont
  {Browaeys}}\ and\ \bibinfo {author} {\bibfnamefont {T.}~\bibnamefont
  {Lahaye}},\ }\bibfield  {title} {\bibinfo {title} {Many-body physics with
  individually controlled rydberg atoms},\ }\href
  {https://doi.org/10.1038/s41567-019-0733-z} {\bibfield  {journal} {\bibinfo
  {journal} {Nature Physics}\ }\textbf {\bibinfo {volume} {16}},\ \bibinfo
  {pages} {132} (\bibinfo {year} {2020})}\BibitemShut {NoStop}%
\bibitem [{Note2()}]{Note2}%
  \BibitemOpen
  \bibinfo {note} {More precisely, the number of intersections is only
  conserved up to a quantity of order $\sim h/J$.}\BibitemShut {Stop}%
\bibitem [{Note3()}]{Note3}%
  \BibitemOpen
  \bibinfo {note} {The exponential scaling is for perturbations more short
  range than some critical power law \cite {Ho, Nayak1, Nayak2}. For
  perturbations of longer range, the scaling of the prethermal timescale may be
  modified. Our arguments should remain valid up to the prethermal timescale,
  whatever it is.}\BibitemShut {Stop}%
\bibitem [{\citenamefont {Pauling}(1935)}]{Pauling1935entropy}%
  \BibitemOpen
  \bibfield  {author} {\bibinfo {author} {\bibfnamefont {L.}~\bibnamefont
  {Pauling}},\ }\bibfield  {title} {\bibinfo {title} {The structure and entropy
  of ice and of other crystals with some randomness of atomic arrangement},\
  }\href@noop {} {\bibfield  {journal} {\bibinfo  {journal} {Journal of the
  American Chemical Society}\ }\textbf {\bibinfo {volume} {57}},\ \bibinfo
  {pages} {2680} (\bibinfo {year} {1935})}\BibitemShut {NoStop}%
\bibitem [{sup()}]{supplement}%
  \BibitemOpen
  \href@noop {} {}\bibinfo {note} {See Supplemental Material (appended) for a
  discussion of (i) the augmented duality mapping, (ii) a discussion of the
  model's relaxation towards equilibrium, and (iii) a discussion of the
  Schrieffer-Wolff transformation to the rotated basis in which the scars are
  maximally simple, and of the nontrivial dynamics arising in the `bare'
  basis}\BibitemShut {NoStop}%
\bibitem [{\citenamefont {Lesanovsky}\ and\ \citenamefont
  {Katsura}(2012)}]{LesanovskyPXP}%
  \BibitemOpen
  \bibfield  {author} {\bibinfo {author} {\bibfnamefont {I.}~\bibnamefont
  {Lesanovsky}}\ and\ \bibinfo {author} {\bibfnamefont {H.}~\bibnamefont
  {Katsura}},\ }\bibfield  {title} {\bibinfo {title} {Interacting fibonacci
  anyons in a rydberg gas},\ }\href
  {https://doi.org/10.1103/PhysRevA.86.041601} {\bibfield  {journal} {\bibinfo
  {journal} {Phys. Rev. A}\ }\textbf {\bibinfo {volume} {86}},\ \bibinfo
  {pages} {041601} (\bibinfo {year} {2012})}\BibitemShut {NoStop}%
\bibitem [{\citenamefont {Turner}\ \emph
  {et~al.}(2018{\natexlab{b}})\citenamefont {Turner}, \citenamefont
  {Michailidis}, \citenamefont {Abanin}, \citenamefont {Serbyn},\ and\
  \citenamefont {Papi\ifmmode~\acute{c}\else \'{c}\fi{}}}]{Turner2018scarred}%
  \BibitemOpen
  \bibfield  {author} {\bibinfo {author} {\bibfnamefont {C.~J.}\ \bibnamefont
  {Turner}}, \bibinfo {author} {\bibfnamefont {A.~A.}\ \bibnamefont
  {Michailidis}}, \bibinfo {author} {\bibfnamefont {D.~A.}\ \bibnamefont
  {Abanin}}, \bibinfo {author} {\bibfnamefont {M.}~\bibnamefont {Serbyn}},\
  and\ \bibinfo {author} {\bibfnamefont {Z.}~\bibnamefont
  {Papi\ifmmode~\acute{c}\else \'{c}\fi{}}},\ }\bibfield  {title} {\bibinfo
  {title} {Quantum scarred eigenstates in a rydberg atom chain: Entanglement,
  breakdown of thermalization, and stability to perturbations},\ }\href
  {https://doi.org/10.1103/PhysRevB.98.155134} {\bibfield  {journal} {\bibinfo
  {journal} {Phys. Rev. B}\ }\textbf {\bibinfo {volume} {98}},\ \bibinfo
  {pages} {155134} (\bibinfo {year} {2018}{\natexlab{b}})}\BibitemShut
  {NoStop}%
\bibitem [{\citenamefont {Michailidis}\ \emph {et~al.}(2020)\citenamefont
  {Michailidis}, \citenamefont {Turner}, \citenamefont
  {Papi\ifmmode~\acute{c}\else \'{c}\fi{}}, \citenamefont {Abanin},\ and\
  \citenamefont {Serbyn}}]{Michailidis2dPXP}%
  \BibitemOpen
  \bibfield  {author} {\bibinfo {author} {\bibfnamefont {A.~A.}\ \bibnamefont
  {Michailidis}}, \bibinfo {author} {\bibfnamefont {C.~J.}\ \bibnamefont
  {Turner}}, \bibinfo {author} {\bibfnamefont {Z.}~\bibnamefont
  {Papi\ifmmode~\acute{c}\else \'{c}\fi{}}}, \bibinfo {author} {\bibfnamefont
  {D.~A.}\ \bibnamefont {Abanin}},\ and\ \bibinfo {author} {\bibfnamefont
  {M.}~\bibnamefont {Serbyn}},\ }\bibfield  {title} {\bibinfo {title}
  {Stabilizing two-dimensional quantum scars by deformation and
  synchronization},\ }\href {https://doi.org/10.1103/PhysRevResearch.2.022065}
  {\bibfield  {journal} {\bibinfo  {journal} {Phys. Rev. Research}\ }\textbf
  {\bibinfo {volume} {2}},\ \bibinfo {pages} {022065} (\bibinfo {year}
  {2020})}\BibitemShut {NoStop}%
\bibitem [{\citenamefont {Lin}\ \emph {et~al.}(2020)\citenamefont {Lin},
  \citenamefont {Calvera},\ and\ \citenamefont {Hsieh}}]{Lin2dPXP}%
  \BibitemOpen
  \bibfield  {author} {\bibinfo {author} {\bibfnamefont {C.-J.}\ \bibnamefont
  {Lin}}, \bibinfo {author} {\bibfnamefont {V.}~\bibnamefont {Calvera}},\ and\
  \bibinfo {author} {\bibfnamefont {T.~H.}\ \bibnamefont {Hsieh}},\ }\bibfield
  {title} {\bibinfo {title} {Quantum many-body scar states in two-dimensional
  rydberg atom arrays},\ }\href {https://doi.org/10.1103/PhysRevB.101.220304}
  {\bibfield  {journal} {\bibinfo  {journal} {Phys. Rev. B}\ }\textbf {\bibinfo
  {volume} {101}},\ \bibinfo {pages} {220304} (\bibinfo {year}
  {2020})}\BibitemShut {NoStop}%
\bibitem [{\citenamefont {Celi}\ \emph {et~al.}(2020)\citenamefont {Celi},
  \citenamefont {Vermersch}, \citenamefont {Viyuela}, \citenamefont {Pichler},
  \citenamefont {Lukin},\ and\ \citenamefont {Zoller}}]{Celi2020}%
  \BibitemOpen
  \bibfield  {author} {\bibinfo {author} {\bibfnamefont {A.}~\bibnamefont
  {Celi}}, \bibinfo {author} {\bibfnamefont {B.}~\bibnamefont {Vermersch}},
  \bibinfo {author} {\bibfnamefont {O.}~\bibnamefont {Viyuela}}, \bibinfo
  {author} {\bibfnamefont {H.}~\bibnamefont {Pichler}}, \bibinfo {author}
  {\bibfnamefont {M.~D.}\ \bibnamefont {Lukin}},\ and\ \bibinfo {author}
  {\bibfnamefont {P.}~\bibnamefont {Zoller}},\ }\bibfield  {title} {\bibinfo
  {title} {Emerging two-dimensional gauge theories in rydberg configurable
  arrays},\ }\href {https://doi.org/10.1103/PhysRevX.10.021057} {\bibfield
  {journal} {\bibinfo  {journal} {Phys. Rev. X}\ }\textbf {\bibinfo {volume}
  {10}},\ \bibinfo {pages} {021057} (\bibinfo {year} {2020})}\BibitemShut
  {NoStop}%
\bibitem [{Note4()}]{Note4}%
  \BibitemOpen
  \bibinfo {note} {Although the lowest-order Hamiltonian is reproduced in
  Ref.~\cite {Celi2020}, the robustness to higher-orders which we describe
  shortly will presumably be absent}\BibitemShut {NoStop}%
\bibitem [{\citenamefont {Serbyn}\ \emph {et~al.}(2021)\citenamefont {Serbyn},
  \citenamefont {Abanin},\ and\ \citenamefont {Papi{\'c}}}]{serbyn2021review}%
  \BibitemOpen
  \bibfield  {author} {\bibinfo {author} {\bibfnamefont {M.}~\bibnamefont
  {Serbyn}}, \bibinfo {author} {\bibfnamefont {D.~A.}\ \bibnamefont {Abanin}},\
  and\ \bibinfo {author} {\bibfnamefont {Z.}~\bibnamefont {Papi{\'c}}},\
  }\bibfield  {title} {\bibinfo {title} {Quantum many-body scars and weak
  breaking of ergodicity},\ }\href {https://doi.org/10.1038/s41567-021-01230-2}
  {\bibfield  {journal} {\bibinfo  {journal} {Nature Physics}\ }\textbf
  {\bibinfo {volume} {17}},\ \bibinfo {pages} {675} (\bibinfo {year}
  {2021})}\BibitemShut {NoStop}%
\bibitem [{\citenamefont {Papi{\'c}}(2021)}]{papic2021review}%
  \BibitemOpen
  \bibfield  {author} {\bibinfo {author} {\bibfnamefont {Z.}~\bibnamefont
  {Papi{\'c}}},\ }\bibfield  {title} {\bibinfo {title} {Weak ergodicity
  breaking through the lens of quantum entanglement},\ }\bibfield  {journal}
  {\bibinfo  {journal} {arXiv preprint arXiv:2108.03460}\ }\href
  {https://doi.org/10.48550/arXiv.2108.03460} {10.48550/arXiv.2108.03460}
  (\bibinfo {year} {2021})\BibitemShut {NoStop}%
\bibitem [{\citenamefont {Moudgalya}\ \emph
  {et~al.}(2022{\natexlab{b}})\citenamefont {Moudgalya}, \citenamefont
  {Bernevig},\ and\ \citenamefont {Regnault}}]{Moudgalya2022review}%
  \BibitemOpen
  \bibfield  {author} {\bibinfo {author} {\bibfnamefont {S.}~\bibnamefont
  {Moudgalya}}, \bibinfo {author} {\bibfnamefont {B.~A.}\ \bibnamefont
  {Bernevig}},\ and\ \bibinfo {author} {\bibfnamefont {N.}~\bibnamefont
  {Regnault}},\ }\bibfield  {title} {\bibinfo {title} {Quantum many-body scars
  and hilbert space fragmentation: a review of exact results},\ }\href
  {https://doi.org/10.1088/1361-6633/ac73a0} {\bibfield  {journal} {\bibinfo
  {journal} {Reports on Progress in Physics}\ }\textbf {\bibinfo {volume}
  {85}},\ \bibinfo {pages} {086501} (\bibinfo {year}
  {2022}{\natexlab{b}})}\BibitemShut {NoStop}%
\bibitem [{Note5()}]{Note5}%
  \BibitemOpen
  \bibinfo {note} {We can furthermore show that the ground state of $H_0$ has
  finite energy density using a variational argument: The state with $|+\rangle
  = \protect \frac {1}{\protect \sqrt {2}}(|0\rangle + |1\rangle )$ states on
  even vertices and $|0\rangle $ states on odd vertices has $\langle \psi
  |\protect \tilde {X}_v|\psi \rangle =1$ for even vertices and $\langle \psi
  |\protect \tilde {X}_v|\psi \rangle =0$ for odd vertices, giving $\langle
  \protect \tilde {X}_v\rangle =\protect \frac {1}{2}$ on average, where
  $\protect \tilde {X}_v = X_v (P^{0000}_v + P^{1111}_v)$ is the conditional
  spin flip operator. Therefore, the true ground state has energy $E<-h
  L^2/2$}\BibitemShut {NoStop}%
\bibitem [{\citenamefont {Qi}\ \emph {et~al.}(2022)\citenamefont {Qi},
  \citenamefont {Hart}, \citenamefont {Friedman}, \citenamefont {Nandkishore},\
  and\ \citenamefont {Lucas}}]{fractonMHD}%
  \BibitemOpen
  \bibfield  {author} {\bibinfo {author} {\bibfnamefont {M.}~\bibnamefont
  {Qi}}, \bibinfo {author} {\bibfnamefont {O.}~\bibnamefont {Hart}}, \bibinfo
  {author} {\bibfnamefont {A.~J.}\ \bibnamefont {Friedman}}, \bibinfo {author}
  {\bibfnamefont {R.}~\bibnamefont {Nandkishore}},\ and\ \bibinfo {author}
  {\bibfnamefont {A.}~\bibnamefont {Lucas}},\ }\href
  {https://doi.org/10.48550/ARXIV.2205.05695} {\bibinfo {title} {Fracton
  magnetohydrodynamics}} (\bibinfo {year} {2022}),\ \Eprint
  {https://arxiv.org/abs/2205.05695} {2205.05695} \BibitemShut {NoStop}%
\bibitem [{\citenamefont {Iaconis}\ \emph {et~al.}(2019)\citenamefont
  {Iaconis}, \citenamefont {Vijay},\ and\ \citenamefont {Nandkishore}}]{IVN}%
  \BibitemOpen
  \bibfield  {author} {\bibinfo {author} {\bibfnamefont {J.}~\bibnamefont
  {Iaconis}}, \bibinfo {author} {\bibfnamefont {S.}~\bibnamefont {Vijay}},\
  and\ \bibinfo {author} {\bibfnamefont {R.}~\bibnamefont {Nandkishore}},\
  }\bibfield  {title} {\bibinfo {title} {Anomalous subdiffusion from subsystem
  symmetries},\ }\href {https://doi.org/10.1103/PhysRevB.100.214301} {\bibfield
   {journal} {\bibinfo  {journal} {Physical Review B}\ }\textbf {\bibinfo
  {volume} {100}},\ \bibinfo {pages} {214301} (\bibinfo {year}
  {2019})}\BibitemShut {NoStop}%
\bibitem [{Note6()}]{Note6}%
  \BibitemOpen
  \bibinfo {note} {We note in passing that our results are presented strictly
  in the Schrieffer-Wolff transformed basis, and in the `bare' basis there can
  be non-trivial dynamics, see \cite {supplement}}\BibitemShut {NoStop}%
\bibitem [{\citenamefont {Geraedts}\ \emph {et~al.}(2017)\citenamefont
  {Geraedts}, \citenamefont {Bhatt},\ and\ \citenamefont
  {Nandkishore}}]{GeraedtsLIOM}%
  \BibitemOpen
  \bibfield  {author} {\bibinfo {author} {\bibfnamefont {S.~D.}\ \bibnamefont
  {Geraedts}}, \bibinfo {author} {\bibfnamefont {R.~N.}\ \bibnamefont
  {Bhatt}},\ and\ \bibinfo {author} {\bibfnamefont {R.}~\bibnamefont
  {Nandkishore}},\ }\bibfield  {title} {\bibinfo {title} {Emergent local
  integrals of motion without a complete set of localized eigenstates},\ }\href
  {https://doi.org/10.1103/PhysRevB.95.064204} {\bibfield  {journal} {\bibinfo
  {journal} {Phys. Rev. B}\ }\textbf {\bibinfo {volume} {95}},\ \bibinfo
  {pages} {064204} (\bibinfo {year} {2017})}\BibitemShut {NoStop}%
\bibitem [{\citenamefont {Ho}\ \emph {et~al.}(2018)\citenamefont {Ho},
  \citenamefont {Protopopov},\ and\ \citenamefont {Abanin}}]{Ho}%
  \BibitemOpen
  \bibfield  {author} {\bibinfo {author} {\bibfnamefont {W.~W.}\ \bibnamefont
  {Ho}}, \bibinfo {author} {\bibfnamefont {I.}~\bibnamefont {Protopopov}},\
  and\ \bibinfo {author} {\bibfnamefont {D.~A.}\ \bibnamefont {Abanin}},\
  }\bibfield  {title} {\bibinfo {title} {Bounds on energy absorption and
  prethermalization in quantum systems with long-range interactions},\ }\href
  {https://doi.org/10.1103/PhysRevLett.120.200601} {\bibfield  {journal}
  {\bibinfo  {journal} {Phys. Rev. Lett.}\ }\textbf {\bibinfo {volume} {120}},\
  \bibinfo {pages} {200601} (\bibinfo {year} {2018})}\BibitemShut {NoStop}%
\bibitem [{\citenamefont {Machado}\ \emph {et~al.}(2019)\citenamefont
  {Machado}, \citenamefont {Kahanamoku-Meyer}, \citenamefont {Else},
  \citenamefont {Nayak},\ and\ \citenamefont {Yao}}]{Nayak1}%
  \BibitemOpen
  \bibfield  {author} {\bibinfo {author} {\bibfnamefont {F.}~\bibnamefont
  {Machado}}, \bibinfo {author} {\bibfnamefont {G.~D.}\ \bibnamefont
  {Kahanamoku-Meyer}}, \bibinfo {author} {\bibfnamefont {D.~V.}\ \bibnamefont
  {Else}}, \bibinfo {author} {\bibfnamefont {C.}~\bibnamefont {Nayak}},\ and\
  \bibinfo {author} {\bibfnamefont {N.~Y.}\ \bibnamefont {Yao}},\ }\bibfield
  {title} {\bibinfo {title} {Exponentially slow heating in short and long-range
  interacting floquet systems},\ }\href
  {https://doi.org/10.1103/PhysRevResearch.1.033202} {\bibfield  {journal}
  {\bibinfo  {journal} {Phys. Rev. Research}\ }\textbf {\bibinfo {volume}
  {1}},\ \bibinfo {pages} {033202} (\bibinfo {year} {2019})}\BibitemShut
  {NoStop}%
\bibitem [{\citenamefont {Machado}\ \emph {et~al.}(2020)\citenamefont
  {Machado}, \citenamefont {Else}, \citenamefont {Kahanamoku-Meyer},
  \citenamefont {Nayak},\ and\ \citenamefont {Yao}}]{Nayak2}%
  \BibitemOpen
  \bibfield  {author} {\bibinfo {author} {\bibfnamefont {F.}~\bibnamefont
  {Machado}}, \bibinfo {author} {\bibfnamefont {D.~V.}\ \bibnamefont {Else}},
  \bibinfo {author} {\bibfnamefont {G.~D.}\ \bibnamefont {Kahanamoku-Meyer}},
  \bibinfo {author} {\bibfnamefont {C.}~\bibnamefont {Nayak}},\ and\ \bibinfo
  {author} {\bibfnamefont {N.~Y.}\ \bibnamefont {Yao}},\ }\bibfield  {title}
  {\bibinfo {title} {Long-range prethermal phases of nonequilibrium matter},\
  }\href {https://doi.org/10.1103/PhysRevX.10.011043} {\bibfield  {journal}
  {\bibinfo  {journal} {Phys. Rev. X}\ }\textbf {\bibinfo {volume} {10}},\
  \bibinfo {pages} {011043} (\bibinfo {year} {2020})}\BibitemShut {NoStop}%
\bibitem [{\citenamefont {Chen}\ \emph {et~al.}(2020)\citenamefont {Chen},
  \citenamefont {Nandkishore},\ and\ \citenamefont
  {Lucas}}]{Chen2020butterfly}%
  \BibitemOpen
  \bibfield  {author} {\bibinfo {author} {\bibfnamefont {X.}~\bibnamefont
  {Chen}}, \bibinfo {author} {\bibfnamefont {R.~M.}\ \bibnamefont
  {Nandkishore}},\ and\ \bibinfo {author} {\bibfnamefont {A.}~\bibnamefont
  {Lucas}},\ }\bibfield  {title} {\bibinfo {title} {Quantum butterfly effect in
  polarized floquet systems},\ }\href
  {https://doi.org/10.1103/PhysRevB.101.064307} {\bibfield  {journal} {\bibinfo
   {journal} {Phys. Rev. B}\ }\textbf {\bibinfo {volume} {101}},\ \bibinfo
  {pages} {064307} (\bibinfo {year} {2020})}\BibitemShut {NoStop}%
\bibitem [{\citenamefont {Feldmeier}\ \emph {et~al.}(2020)\citenamefont
  {Feldmeier}, \citenamefont {Sala}, \citenamefont {De~Tomasi}, \citenamefont
  {Pollmann},\ and\ \citenamefont {Knap}}]{FeldmeierDipole}%
  \BibitemOpen
  \bibfield  {author} {\bibinfo {author} {\bibfnamefont {J.}~\bibnamefont
  {Feldmeier}}, \bibinfo {author} {\bibfnamefont {P.}~\bibnamefont {Sala}},
  \bibinfo {author} {\bibfnamefont {G.}~\bibnamefont {De~Tomasi}}, \bibinfo
  {author} {\bibfnamefont {F.}~\bibnamefont {Pollmann}},\ and\ \bibinfo
  {author} {\bibfnamefont {M.}~\bibnamefont {Knap}},\ }\bibfield  {title}
  {\bibinfo {title} {Anomalous diffusion in dipole- and
  higher-moment-conserving systems},\ }\href
  {https://doi.org/10.1103/PhysRevLett.125.245303} {\bibfield  {journal}
  {\bibinfo  {journal} {Phys. Rev. Lett.}\ }\textbf {\bibinfo {volume} {125}},\
  \bibinfo {pages} {245303} (\bibinfo {year} {2020})}\BibitemShut {NoStop}%
\bibitem [{\citenamefont {Iaconis}\ \emph {et~al.}(2021)\citenamefont
  {Iaconis}, \citenamefont {Lucas},\ and\ \citenamefont
  {Nandkishore}}]{IaconisAnyDimension}%
  \BibitemOpen
  \bibfield  {author} {\bibinfo {author} {\bibfnamefont {J.}~\bibnamefont
  {Iaconis}}, \bibinfo {author} {\bibfnamefont {A.}~\bibnamefont {Lucas}},\
  and\ \bibinfo {author} {\bibfnamefont {R.}~\bibnamefont {Nandkishore}},\
  }\bibfield  {title} {\bibinfo {title} {Multipole conservation laws and
  subdiffusion in any dimension},\ }\href
  {https://doi.org/10.1103/PhysRevE.103.022142} {\bibfield  {journal} {\bibinfo
   {journal} {Phys. Rev. E}\ }\textbf {\bibinfo {volume} {103}},\ \bibinfo
  {pages} {022142} (\bibinfo {year} {2021})}\BibitemShut {NoStop}%
\bibitem [{\citenamefont {Henley}(2010)}]{HenleyAnnuRev}%
  \BibitemOpen
  \bibfield  {author} {\bibinfo {author} {\bibfnamefont {C.~L.}\ \bibnamefont
  {Henley}},\ }\bibfield  {title} {\bibinfo {title} {The ``coulomb phase'' in
  frustrated systems},\ }\href
  {https://doi.org/10.1146/annurev-conmatphys-070909-104138} {\bibfield
  {journal} {\bibinfo  {journal} {Annual Review of Condensed Matter Physics}\
  }\textbf {\bibinfo {volume} {1}},\ \bibinfo {pages} {179} (\bibinfo {year}
  {2010})}\BibitemShut {NoStop}%
\bibitem [{\citenamefont {Castelnovo}\ \emph {et~al.}(2012)\citenamefont
  {Castelnovo}, \citenamefont {Moessner},\ and\ \citenamefont
  {Sondhi}}]{CastelnovoAnnuRev}%
  \BibitemOpen
  \bibfield  {author} {\bibinfo {author} {\bibfnamefont {C.}~\bibnamefont
  {Castelnovo}}, \bibinfo {author} {\bibfnamefont {R.}~\bibnamefont
  {Moessner}},\ and\ \bibinfo {author} {\bibfnamefont {S.}~\bibnamefont
  {Sondhi}},\ }\bibfield  {title} {\bibinfo {title} {Spin ice,
  fractionalization, and topological order},\ }\href
  {https://doi.org/10.1146/annurev-conmatphys-020911-125058} {\bibfield
  {journal} {\bibinfo  {journal} {Annual Review of Condensed Matter Physics}\
  }\textbf {\bibinfo {volume} {3}},\ \bibinfo {pages} {35} (\bibinfo {year}
  {2012})}\BibitemShut {NoStop}%
\bibitem [{Note7()}]{Note7}%
  \BibitemOpen
  \bibinfo {note} {We caution that quantum circuit dynamics may not accurately
  capture the scaling of diffusion constants in the Hamiltonian system at low
  defect density. See, for instance, Ref.~\cite
  {Chen2020butterfly}.}\BibitemShut {Stop}%
\bibitem [{Note8()}]{Note8}%
  \BibitemOpen
  \bibinfo {note} {Note that the random scar states are not perfectly aligned
  along $x$; the typical value of $|\protect \mathcal {Q}_y|$ is set by
  $\protect \sqrt {L}$. This leads to an additional source of diffusive
  broadening in the transverse direction.}\BibitemShut {Stop}%
\end{thebibliography}%
\let\addcontentsline\oldaddcontentsline

\cleardoublepage
\newpage

\onecolumngrid
\begin{center}
\textbf{\large Supplemental Material for ``Ergodicity breaking provably robust to arbitrary perturbations''}
\vskip 0.4cm
{David{\;\,}T.{\;\,}Stephen,\textsuperscript{1,\,2}{\;\,}Oliver{\;\,}Hart,\textsuperscript{1}{\;\,}and{\;\,}Rahul{\;\,}M.{\;\,}Nandkishore\textsuperscript{1}}
\vskip 0.1cm
\textsuperscript{1}{\fontsize{9.5pt}{11.5pt}\selectfont\emph{Department of Physics and Center for Theory of Quantum Matter,\\[-0.75pt] University of Colorado, Boulder, Colorado 80309, USA}}\\[-0.75pt]
\textsuperscript{2}{\fontsize{9.5pt}{11.5pt}\selectfont\emph{Department of Physics, California Institute of Technology, Pasadena, California 91125, USA}}\\[-0.75pt]
{\fontsize{9pt}{11pt}\selectfont(Dated: October 5, 2022)}
\vskip 0.75cm
\end{center}
\twocolumngrid

\setcounter{secnumdepth}{2}

\tableofcontents


\section{Proof of robustness of the scar states} \label{app:robustness_proof}

Here we give a more rigorous argument that the scar states based on dense packings of domain walls are robust to all orders in perturbation theory. Let us consider the operator $X_I = \prod_{i\in I} X_i$ which flips a subset of spins labelled by $i\in I$, such that $|I|<L$. This restriction on the size of $I$ means that it cannot contain all sites along any non-contractible loop around the torus. Therefore, we can contain all of the sites in $I$ within a contractible, ball-shaped region. More precisely, there is a connected, contractible surface on the square lattice defined by a subset of plaquettes such that $X_I$ acts only on spins contained within these plaquettes.

Now, consider the uppermost row in the lattice which intersects $I$, such that the row above contains no sites in $I$. Such a row can always be identified given the contractability assumption defined above. Let $i_\text{NW},i_\text{NE}\in I$ be the leftmost and rightmost elements of $I$ belonging to this row. Define corresponding plaquettes $p_\text{NW}$ and $p_\text{NE}$ as the plaquettes whose lower-right (lower-left) site is $i_\text{NW}$ ($i_\text{NE}$). By definition, the only site belonging to $p_\text{NW}$ ($p_\text{NE}$) that is acted on by $X_{I}$  is $i_\text{NW}$ ($i_\text{NE}$).

We now show that $X_{I}$ flips the eigenvalue of either $\CZ_{p_\text{NW}}$ or $\CZ_{p_\text{NE}}$. Recall that the scars states are defined by the pattern `0011' on every row, and that pattern is shifted by one site to the left or right between neighbouring rows. Suppose that the uppermost row containing sites in $I$ has this pattern shifted to the right compared to the row above it. Then, it is easy to check that acting with $X_{i_\text{NW}}$ results in $\CZ_{p_\text{NW}}=-1$. If the uppermost row is instead shifted to the left, then acting with $X_{i_\text{NE}}$ results in $\CZ_{p_\text{NE}}=-1$. Therefore, $X_I$ is annihilated after projection onto the $\CZ_p=1$ subspace, meaning that no product of spin flips which acts on fewer than $L$ sites can act on any of the scar states.


\section{Size of no-intersections sector}

The local constraint $\CZ_p = 1$ for all plaquettes $p$ defines a restricted Hilbert space $\mathcal{H}_0$ whose dimension grows nontrivially with system volume $L^2$. The dimension of this hard-constrained Hilbert space can be enumerated exactly for sufficiently small systems, allowing us to approximate the asymptotic behavior with $L$. The trivial case of an isolated plaquette, consisting of four spins, was considered in the main text: 12 of the possible 16 spin configurations are compatible with the local constraint. To study the scaling with $L$, we employ a transfer matrix approach. Let us denote the binary representation of the spin at position $\vec{r}$ by $b_{\vec{r}} \in \{ 0, 1 \}$, i.e., $b_\vec{r} = \frac12 (1 - Z_\vec{r})$.
In this notation, the $\CZ$ gate acting on two spins located at positions $\vec{r}_1$ and $\vec{r}_2$ can be written
\begin{equation}
    \CZ(\vec{r}_1, \vec{r}_2) = \exp\left[ i\pi b_{\vec{r}_1} b_{\vec{r}_2} \right]
    \, ,
\end{equation}
which gives a minus sign only if $b_{\vec{r}_1} = b_{\vec{r}_2} = 1$.

\begin{figure}[t]
    \centering
    \includegraphics[width=0.695\linewidth]{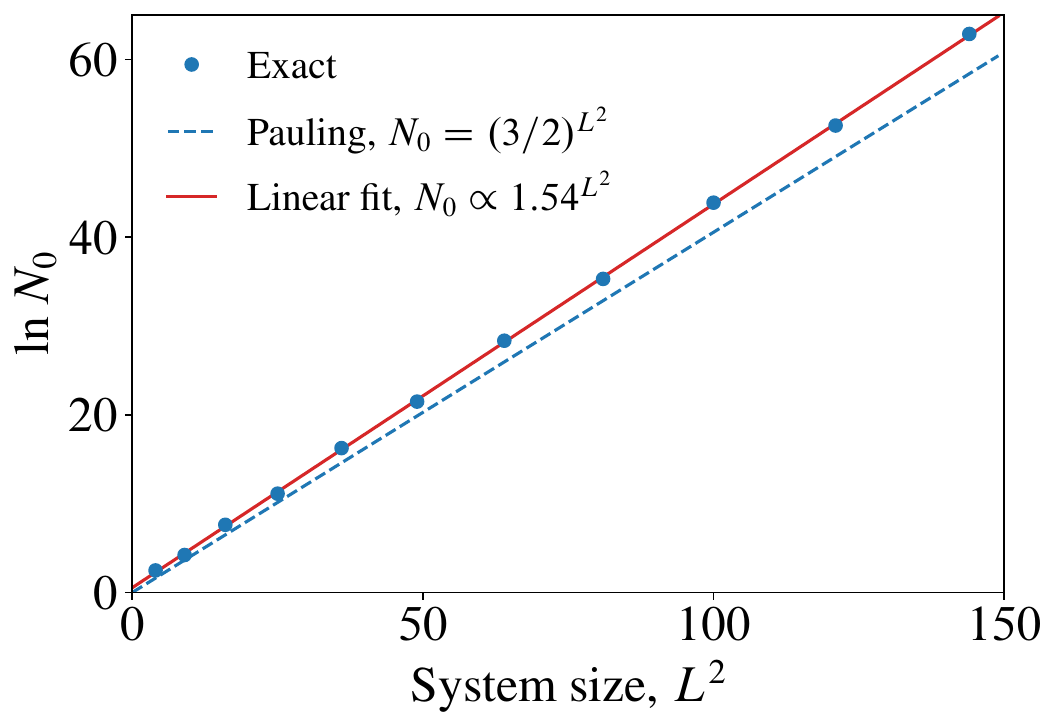}
    \caption{Exact enumeration of the restricted Hilbert space dimension, corresponding to the number of Ising spin configurations that satisfy $\CZ_p = 1$ for all plaquettes $p$. The exact result (blue circles), obtained using exact numerics, is contrasted with the approximate analytical result, discussed in the main text, which treats the local constraints on average (blue dashed line), and the optimal linear fit (red solid line).}
    \label{fig:restricted-dimension}
\end{figure}

Decomposing the two-dimensional system into columns (say) and working from left to right, the transfer matrix tells us which spin configurations in a given column are compatible with the spin configuration of the `previous' column, given the local constraints $\CZ_p = 1$. Disallowed configurations are associated with a vanishing coefficient in the transfer matrix, while allowed transitions are associated with a unit coefficient. This construction gives rise to a transfer matrix whose matrix elements are
\begin{equation}
    T = 
    \prod_{n=1}^L
    \frac12 \left[ 1 + e^{i\pi(i_n i_{n+1} + i_n o_n + i_{n+1} o_{n+1} + o_n o_{n+1})} \right]
\end{equation}
where $\{ i_n \}$ label the `input' bits on one column and $\{ o_n \}$ the `output' bits.
In the presence of periodic boundary conditions, the total number of states that are compatible with the local constraints is recovered by evaluating
\begin{equation}
    \dim \mathcal{H}_0 \equiv N_0 = \Tr T^L
    \, .
    \label{eqn:dimension-CZ1}
\end{equation}
The resulting $N_0$ for systems of size up to and including $L^2 = 144$ spins are shown in Fig.~\ref{fig:restricted-dimension}.
The asymptotic behavior is well described by exponential growth with system size, $N_0 \sim e^{\alpha L^2}$, with $\alpha \approx \ln(1.54)$, which is to be compared with the approximate analytical value $\alpha_\text{P} = \ln(3/2)$ obtained using a Pauling estimate that treats the local constraints on average, as described in the main text.

\section{Refined duality mapping}

In the main text, we described a duality mapping from spins to two species of domain wall variables, corresponding to domain walls between the odd and the even sublattices. This mapping is locally four-to-one, and is sufficient to describe why the scar states are absolutely stable.
However, in order to characterize \emph{all} `intermediate' sectors, which contain a nonvanishing number of non-contractible domain wall loops, we describe here how the duality mapping can be refined from four-to-one to two-to-one by endowing the domain wall variables with an orientation.
While the mapping is still not one-to-one, it is sufficient to describe \emph{all} dynamically disconnected sectors in the computational basis, as we will describe.

\begin{figure}%
    \centering%
    \includegraphics[scale=0.32]{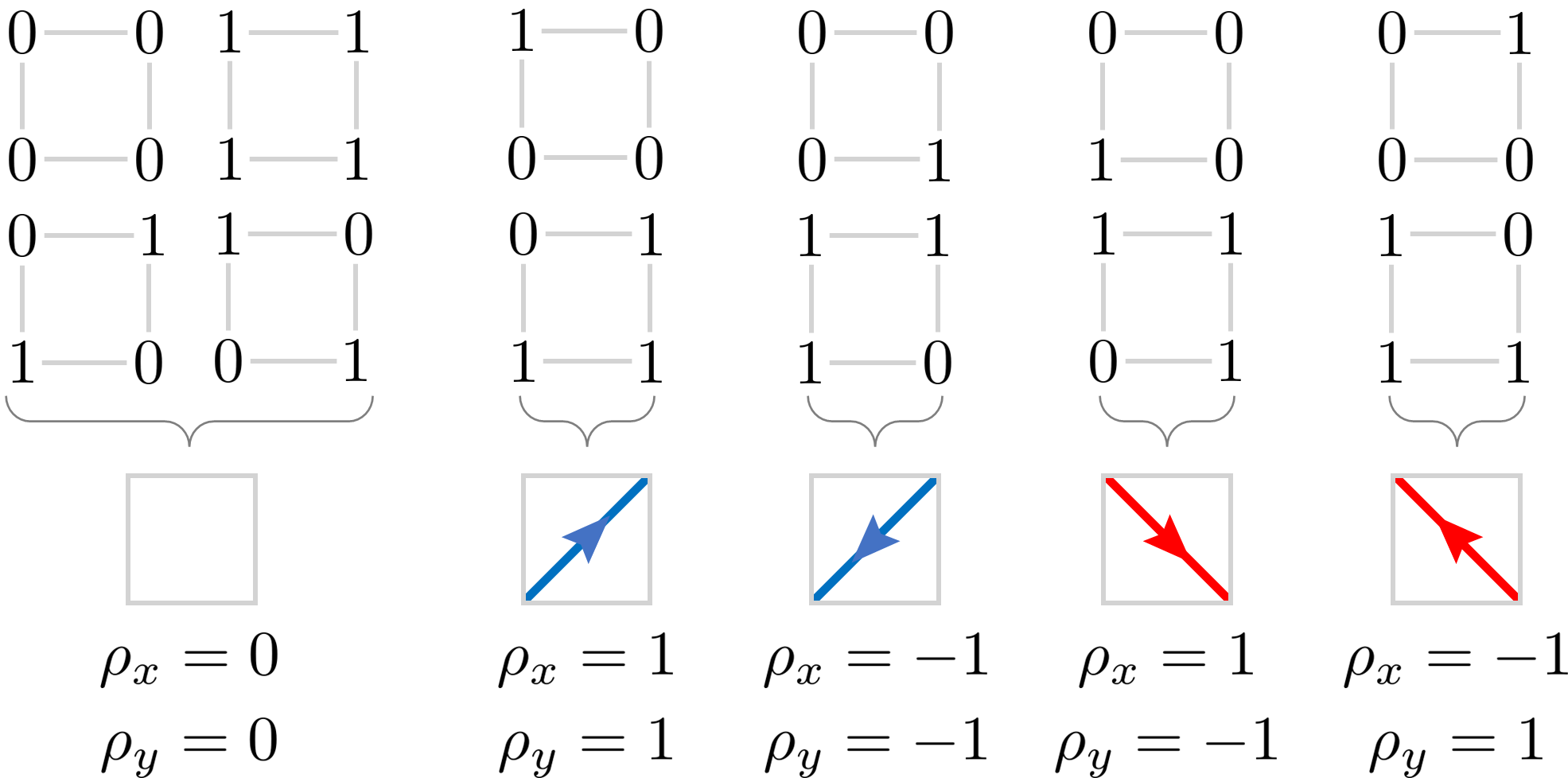}%
    \caption{The refined duality mapping where we retain more information about the original spins in order to define an orientation of domain walls. The depicted mappings are valid for plaquettes on the even sublattice  [i.e., $(-1)^{r_x+r_y}=1$]. The mapping for the odd sublattice is obtained by swapping red and blue and flipping all arrows.}%
    \label{fig:duality_oriented}%
\end{figure}

\begin{figure}
    \centering
    \includegraphics[width=0.325\linewidth]{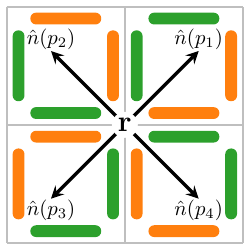}%
    \hspace{1.1cm}%
    \includegraphics[width=0.325\linewidth]{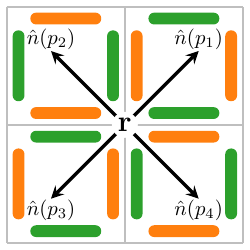}
    \caption{Illustration of the unit vectors $\hat{n}_i(p)$ that appear in the lattice divergence in Eq.~\eqref{eqn:lattice-divergence}. Left: The colors represent the sign that accompanies $Z_i Z_j$ for neighboring sites $i$ and $j$ after summing all contributions to Eq.~\eqref{eqn:lattice-divergence}. Right: These contributions can be regrouped to show that the divergence vanishes as a consequence of Eq.~\eqref{eqn:ZZ-equivalence} for each of the four plaquettes.}
    \label{fig:div-schematic}
\end{figure}

The refined mapping is depicted in Fig.~\ref{fig:duality_oriented}: Each domain wall configuration is assigned a direction according to the values of the local operators
\begin{subequations}
\begin{align}
    \rho_x(\vec{r}) &= \frac{1}{2}  (-1)^{r_x + r_y} ( \tau_{r_x + \frac12, r_y} -  \tau_{r_x - \frac12, r_y} ) \label{eqn:rho-x} \\
    \rho_y(\vec{r}) &= \frac{1}{2}  (-1)^{r_x + r_y} ( \tau_{r_x, r_y-\frac12} -  \tau_{r_x, r_y + \frac12} )
\end{align}%
\label{eqn:conserved-density}%
\end{subequations}
which define the components of a vector $\boldsymbol{\rho}=(\rho_x, \rho_y)$. The vector $\vec{r}=(r_x, r_y)$ here labels the centers of the plaquettes, and we defined the domain wall variables $\tau_{r_x, r_y \pm \frac12} \equiv Z_{r_x + \frac12 , r_y \pm \frac12}Z_{r_x - \frac12 , r_y \pm \frac12}$ and $\tau_{r_x \pm \frac12, r_y} \equiv Z_{r_x \pm \frac12 , r_y + \frac12}Z_{r_x \pm \frac12 , r_y - \frac12}$, which live on the links between lattice sites. In this way, $\rho_x$ and $\rho_y$ depend on the four spins $Z_{r_x \pm \frac12, r_y \pm \frac12}$ surrounding the plaquette at $\vec{r}$, paired parallel to $y$ and $x$, respectively. When the domain wall variables are assigned an orientation according to Eq.~\eqref{eqn:conserved-density}, they form closed loops, which follows from the vanishing of the lattice divergence
\begin{equation}
    (\Div \boldsymbol{\rho})_\vec{r} = \sum_{p \, \cap \, \vec{r}} \hat{n}_i(p)  \rho_i(p)
    \, ,
    \label{eqn:lattice-divergence}
\end{equation}
where the vector index $i$ is implicitly summed over, $\hat{n}_i(p)$ is the unit vector that points from the lattice site at $\vec{r}$ to the center of one of the neighboring plaquettes~(see Fig.~\ref{fig:div-schematic}), and $p \cap \vec{r}$ denotes all plaquettes $p$ that intersect with $\vec{r}$.
For states $\ket{\Psi} \in \mathcal{H}_0$, i.e., belonging to the constrained Hilbert space satisfying $\CZ_p = 1$, we have $(\Div \boldsymbol{\rho})_\vec{r} \ket{\Psi} = 0$ identically for all sites $\vec{r}$, implying that -- once appropriately coarse grained -- $\rho_i$ defines a divergence-free vector field. 
To prove this, let us label the four spins around a single plaquette $p$ in the clockwise direction as $i$, $j$, $k$, and $\ell$. The $\CZ_p$ interaction amongst these four spins can be written in terms of pairwise interactions `across' the plaquette and a four-spin interaction:
\begin{equation}
    \CZ_p = \frac12 \left[ \mathds{1} + Z_i Z_k + Z_j Z_\ell - Z_i Z_j Z_k Z_\ell \right]
    \, .
\end{equation}
This expression implies that, for states belonging to the constrained Hilbert space defined by $\CZ_p=1$, we have
\begin{equation}
    Z_i Z_j + Z_k Z_\ell \cong Z_j Z_k + Z_\ell Z_i \quad \text{for } \ket{\Psi} \in \mathcal{H}_0
    \, .
    \label{eqn:ZZ-equivalence}
\end{equation}
That is, the action of the two operators on either side of Eq.~\eqref{eqn:ZZ-equivalence} is identical when acting on states belonging to $\mathcal{H}_0$.
As shown in Fig.~\ref{fig:div-schematic}, the lattice divergence \eqref{eqn:lattice-divergence} can be grouped into four terms that can be identified with the four plaquettes that intersect $\vec{r}$, and each group vanishes by virtue of the congruence in Eq.~\eqref{eqn:ZZ-equivalence}.

The vanishing of the lattice divergence gives rise to a number of conserved quantities.
Specifically, we are able to identify the following ``one-form'' charges:
\begin{subequations}
\begin{align}
    \mathcal{Q}_x(r_x) &= \sum_{r_y=1}^{L} \rho_x(\vec{r}) \label{eqn:Qx} \\
    \mathcal{Q}_y(r_y) &= \sum_{r_x=1}^{L} \rho_y(\vec{r}) \label{eqn:Qy}
\end{align}%
\label{eqn:one-form-Q}%
\end{subequations}
both of which commute with the effective Hamiltonian $H_0$ given in Eq.~\eqref{eq:effective-ham} in the main text. 
The property~\eqref{eqn:ZZ-equivalence} can also be used to show that $\mathcal{Q}_x(r_x) = \mathcal{Q}_x(r_x+1)$ (and similarly for $\mathcal{Q}_y$),
such that $\mathcal{Q}_x$ and $\mathcal{Q}_y$ do not depend upon the position at which they are evaluated.
Physically, the conserved quantities \eqref{eqn:Qx} and \eqref{eqn:Qy} correspond to the flux of $\rho_i$ through a ``surface'' of codimension one that wraps around the torus in one direction. Since $\mathcal{Q}_x$ and $\mathcal{Q}_y$ are evaluated along a non-contractible path, every contractible domain wall that passes through the surface must also do so in the opposite direction, leading to a vanishing net contribution. On the other hand, if there exists a non-contractible domain wall loop that winds around the torus in the $i$th direction, then this will give rise to a nontrivial value of $\mathcal{Q}_i$ (with domain wall loops that wind in both the $x$ and $y$ directions simultaneously giving nontrivial contributions to both $\mathcal{Q}_x$ and $\mathcal{Q}_y$).
In this way, $|\mathcal{Q}_i|$ counts the number of winding domain walls that span the torus in the $i$th direction (that cannot be removed by local moves, which allow pairwise annihilation of adjacent loops of the same color). The sign of $\mathcal{Q}_i$ characterizes sublattice $\Z2$ symmetry breaking, since Eq.~\eqref{eqn:conserved-density} implies that $X_\text{odd} \mathcal{Q}_i X_\text{odd} = -\mathcal{Q}_i$, where $X_\text{odd}$ is the product of $X_i$ operators on all sites $i$ belonging to the odd sublattice, and similarly for the even sublattice. As described in the main text, if there are winding domain wall loops present, i.e., $|\mathcal{Q}_i| > 0$, then it is possible to connect states that differ by flipping all spins, but it is not possible to connect states that differ by flipping all spins belonging to just one sublattice. In the $\mathcal{Q}_i=0$ sector, the even and odd sublattices can be flipped separately under local dynamics.

\begin{figure}
    \centering
    \includegraphics[width=0.55\linewidth]{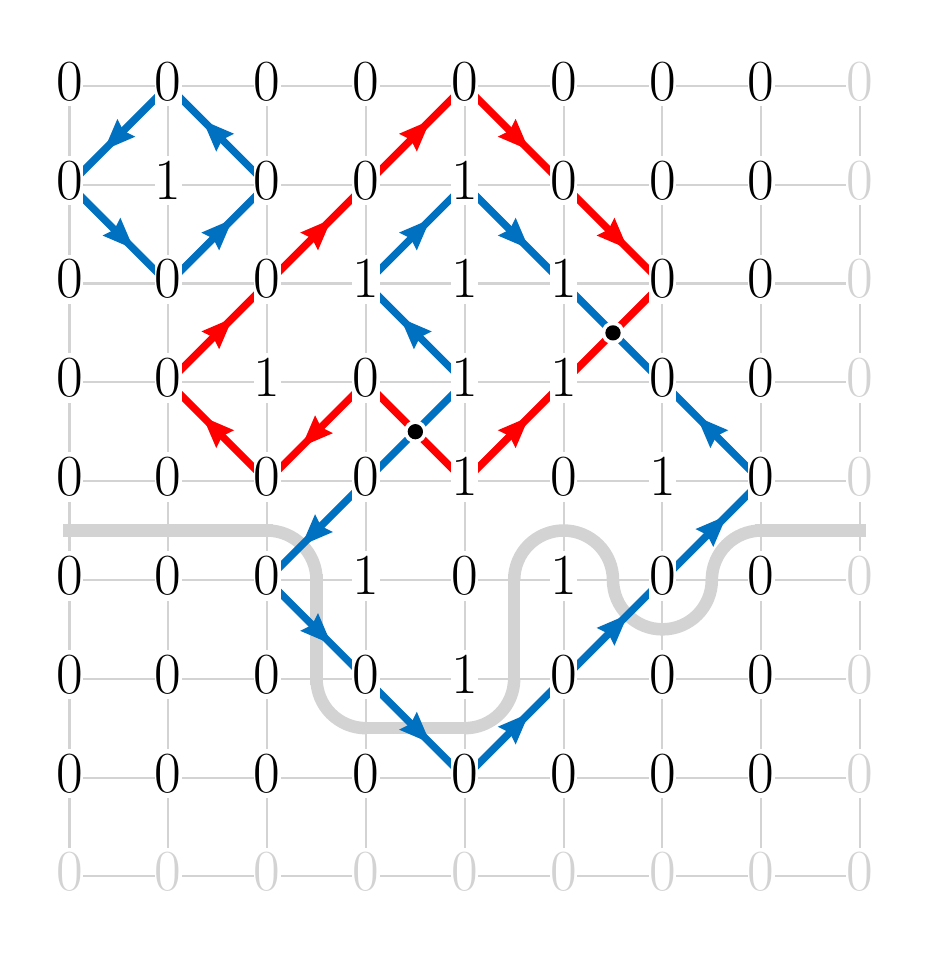}
    \caption{Spin configuration from Fig.~1(b) in the main text labelled according to the refined duality mapping that provides the domain walls with an orientation. When red and blue domain walls intersect, one crossing acts as a `source' of flux, while the other crossing acts as a `sink'. Also shown is an example of a deformed path, along which a one-form charge can be defined, as in Eq.~\eqref{eqn:one-form-Q}.}
    \label{fig:sources-and-sinks}
\end{figure}

If one wishes to locally deform the non-contractible paths in Eq.~\eqref{eqn:one-form-Q} away from being perfectly straight, then they should still pass through the centers of the links (i.e., connecting the centers of the plaquettes), as depicted in Fig.~\ref{fig:sources-and-sinks}. Once again, Eq.~\eqref{eqn:ZZ-equivalence} can be used to find the appropriate linear combinations of $\rho_x$ and $\rho_y$ [i.e., the two choices of sign in $\frac12(\pm \rho_x\pm\rho_y)$] that should be taken on the plaquettes that correspond to a `corner', so as to count whether or not a domain wall has passed through the path (with the appropriate sign).

\section{Numerical analysis of Krylov sectors}

Here we provide some additional details relating to the exact enumeration of Krylov sectors presented in the main text.
To identify the disconnected sub-graphs of the \emph{effective} Hamiltonian, we make use of a breadth first search (BFS) of the system's adjacency matrix.
Since the adjacency matrix is sparse, we need only store at most $O( N_0 L^2 )$ connections between states, where $N_0$ is defined in Eq.~\eqref{eqn:dimension-CZ1}. 
The classification of disconnected sectors in the computational basis proceeds as follows.
We keep track of whether each state belonging to the restricted Hilbert space $\mathcal{H}_0$ has been visited by the algorithm.
In a loop over all states belonging to $\mathcal{H}_0$, if the state has not yet been visited, then it acts as the root node for a new BFS. All neighbors of the root node are added to a queue. For all states in the queue, their neighbors are added to the queue if they have not yet been visited, and the state is subsequently dequeued. This procedure is repeated until the queue is empty, at which point all states that can be reached from the root node have been classified and added to a Krylov sector, and the BFS is complete. The loop over all states belonging to the restricted Hilbert space ensures that all physical states are classified as belonging to a unique Krylov sector once the algorithm is complete.

\begin{figure}[t]
    \centering
    \includegraphics[height=0.6\linewidth]{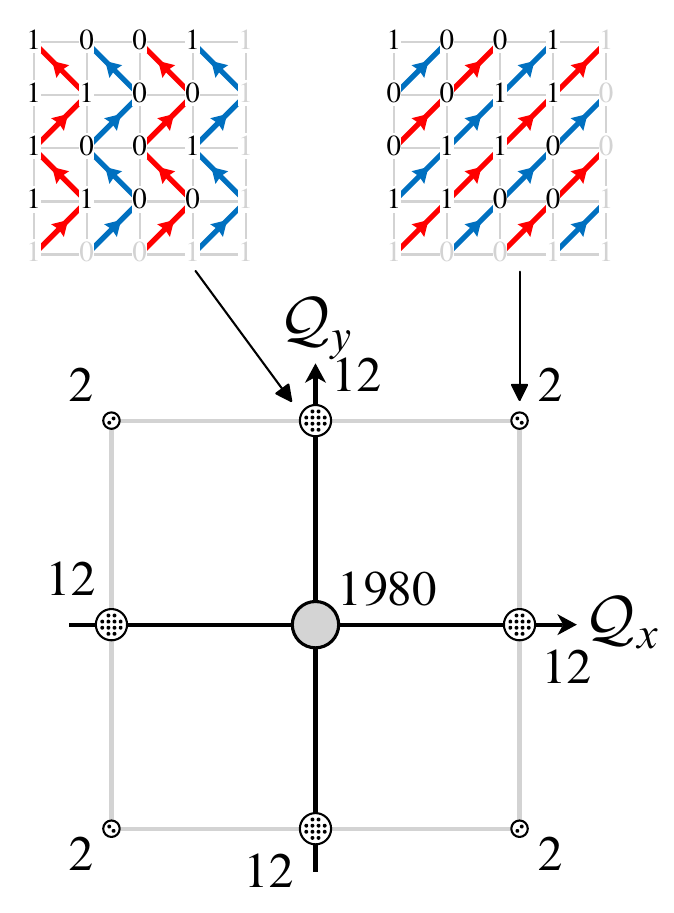}%
    \hfill%
    \includegraphics[height=0.6\linewidth]{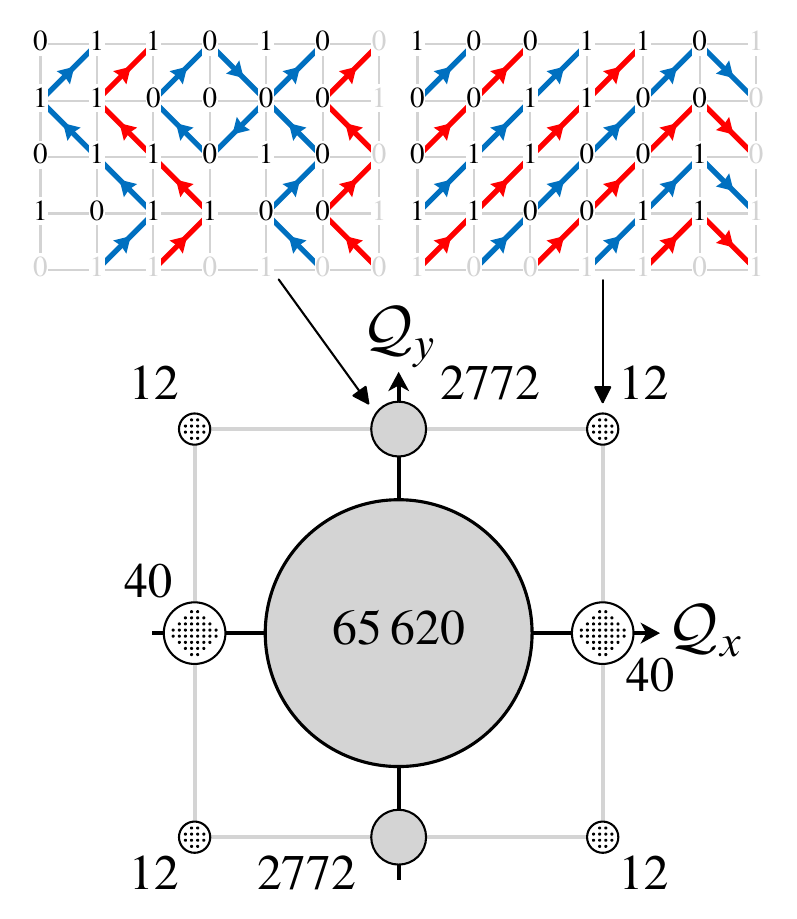}
    \caption{Illustration of symmetry sectors labelled by the one-form symmetry quantum numbers $\mathcal{Q}_x$ and $\mathcal{Q}_y$ [see Eq.~\eqref{eqn:one-form-Q}] for a system of size $(L_x, L_y) = (4, 4)$ (left) and $(6, 4)$ (right). The numbers correspond to the number of states in each symmetry sector, and $\mathcal{Q}_i \in \{ -4, 0, 4 \}$. Non-shaded symmetry sectors fragment further into isolated sectors indicated by small dots and host the scars discussed in the main text. Above are visualizations of some representative states belonging to the symmetry sectors indicated by the arrows.}
    \label{fig:subsectors-zbasis}
\end{figure}

This analysis of sectors is applied to systems of size $(L_x,L_y)=(4,4)$ and $(6, 4)$ in Fig.~\ref{fig:subsectors-zbasis}, resolved by the quantum numbers $\mathcal{Q}_x$ and $\mathcal{Q}_y$ defined in Eq.~\eqref{eqn:one-form-Q}, which label the ``flux'' of (oriented) domain walls through a non-contractible loop of the torus.
For $(L_x,L_y)=(4,4)$, there are $56$ scars in total, and all scar states maximize at least one of $|\mathcal{Q}_x|$ or $|\mathcal{Q}_y|$ (a general property).
The largest sector, which includes the domain wall vacuum, occurs at $(\mathcal{Q}_x, \mathcal{Q}_y) = (0, 0)$.
The case $(L_x,L_y)=(6,4)$, shown in the right panel, illustrates the existence of intermediate Krylov sectors that can occur when the packing of winding domain wall loops is not dense.
Since $L_x$ is not divisible by $4$, it is not possible to write down a spin configuration that gives rise to a dense alternating pattern of red and blue domain walls that all wrap around the torus in the $y$ direction with the same orientation. As a result, there will be some ``empty space,'' corresponding to the domain wall vacuum, allowing the domain walls to move about. The sectors at $(\mathcal{Q}_x, \mathcal{Q}_y) = (0, \pm 4)$ are therefore not further fragmented. Since $L_y$ \emph{is} divisible by 4, there are, however, scars that maximize $|\mathcal{Q}_x|$, which wrap around the torus parallel to $x$. For $(L_x, L_y) = (6, 6)$ (data not shown), such that neither $L_x$ nor $L_y$ is divisible by 4, states that correspond to a dense packing of domain walls cannot be written down and there are no scars.


\begin{figure*}[t]
    \centering
    \includegraphics[width=0.75\linewidth]{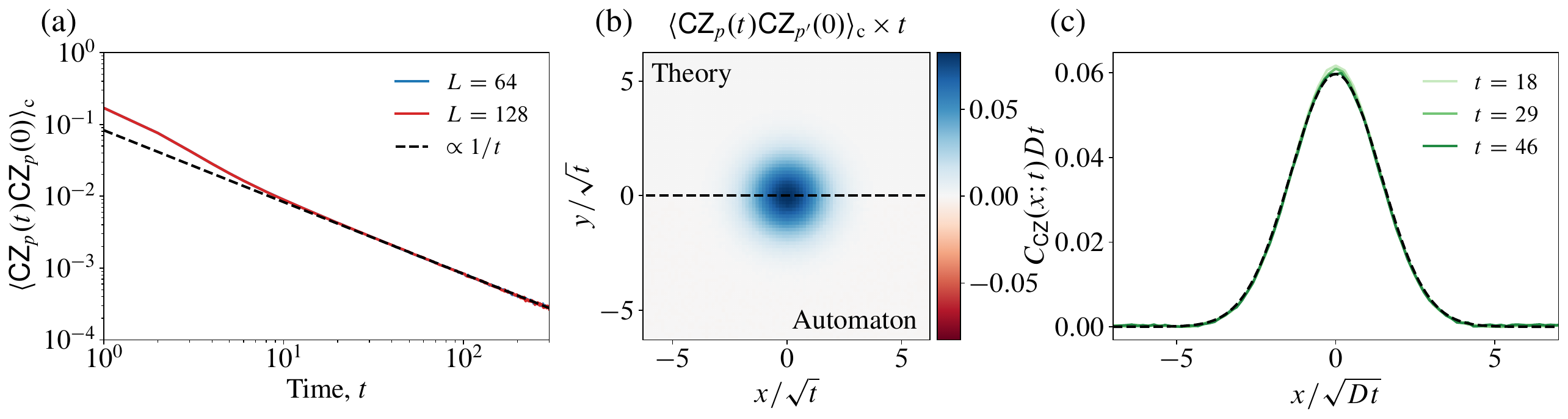}
    \caption{(a) Decay of the plaquette autocorrelation function obtained from the stochastic automaton circuit (solid lines), which asymptotically exhibits diffusive behavior, $t^{-1}$ (dashed black line). (b) Profile of the plaquette correlation function $C_{\mathsf{CZ}}(\vec{r}_{pp'}; t)$ [see Eq.~\eqref{eqn:Cp-definition}] obtained from the stochastic automaton circuit. The numerical results (lower panel) are compared with the Gaussian prediction of Eq.~\eqref{eqn:Cp-analyitcal} (upper panel). Results are for a system of size $L=256$, plotted at time $t \approx 50$. (c) Scaling collapse of the profile along the line $y=0$. In all panels, the stochastic circuit was averaged over at least $10^{5}$ histories.}
    \label{fig:crossing-diffusion}
\end{figure*}

\section{Automaton circuits}
\label{sec:automaton}

To study the behavior of correlation functions such as $\langle Z_i(t) Z_j(0) \rangle$, where the average is evaluated with respect to some ensemble that will be specified in each case, we introduce a stochastic cellular automaton time evolution~\cite{IVN, FeldmeierDipole}, which proceeds as follows.
Each gate that comprises the automaton circuit maps a (product state) configuration of spins in the `computational basis' $\ket{\sigma_i^z}$ to another product state in the computational basis.
In each `step' a candidate spin is chosen from the lattice at random. If flipping the spin does not change $\sum_p \CZ_p$, then the spin is flipped, else it remains unaffected by the automaton gate.
Since this process only (potentially) modifies the value of $\CZ_p$ on the adjacent four plaquettes that overlap with the candidate spin, one can define
a local gate of size $3 \times 3$, centered on the candidate spin, that effects the dynamics.
In the $\CZ_p = 1$ sector, the central spin can only flip if \emph{all} neighboring spins are in the state `0', or if they are all in the state `1'.
Outside of the $\CZ_p = 1$ sector, the update rule also depends on the configuration of the next-nearest-neighbor spins.
The `single spin flip' update is repeated $L^2$ times to propagate the system forwards by one unit of time.
While we choose to work with a random circuit geometry, one can alternatively consider a `brickwork'-style circuit architecture, as in, e.g., Refs.~\cite{IVN,IaconisAnyDimension}.
Unless otherwise stated, time is measured in units of $h^{-1}$ throughout this section.


\subsection{Infinite temperature: Dynamics of crossings}
\label{sec:crossing-diffusion}

First, we consider ``infinite temperature'' dynamics, where temperature is taken to be much larger than both microscopic energy scales $J$ and $h$ appearing in the Hamiltonian. As a result, all product state spin configurations in the computational basis are sampled with equal weight, meaning that the $\CZ_p=1$ constraint is violated in almost all states.
Since the effective (projected) Hamiltonian has an exact $\U1$ conservation law, which corresponds to conservation of the number of domain wall crossings, we expect that such crossings should diffuse asymptotically.
Generically, if higher-order terms in perturbation theory are taken into account, the number of bare crossings is conserved up to a quantity of order $\sim h/J$, up to the prethermal time scale~\cite{ADHH}.
Physically, this conservation law should manifest in the automaton circuit as a slow, $1/t$ decay in the plaquette autocorrelation function
\begin{equation}
    \langle \CZ_p(t) \CZ_p(0) \rangle - \langle \CZ_p(t) \rangle  \langle \CZ_p(0) \rangle \sim 1/t^{d/2}
    \, .
    \label{eqn:CZ-auto}
\end{equation}
Note that the infinite temperature average $\langle \CZ_p(0) \rangle$ does not vanish: $\langle \CZ_p(0) \rangle = (+1) \times 3/4 + (-1) \times 1/4 = 1/2$, since only $4/16$ of the local plaquette configurations host a domain wall crossing.
The autocorrelation function~\eqref{eqn:CZ-auto} is shown in Fig.~\ref{fig:crossing-diffusion}\hyperref[fig:crossing-diffusion]{(a)} for various system sizes. In all cases, over the times plotted, the correlation function exhibits excellent agreement with the diffusive $t^{-1}$ prediction at sufficiently late times ($t \gtrsim 10$).

Additionally, we can characterize the spatial profile of correlations at a fixed but large time $t \gg 1$. In this case, Eq.~\eqref{eqn:CZ-auto} is generalized to 
\begin{align}
    C_\mathsf{CZ} (\vec{r}_{pp'}; t) &\equiv \langle \CZ_p(t) \CZ_{p'}(0) \rangle - \langle \CZ_{p}(t) \rangle  \langle \CZ_{p'}(0) \label{eqn:Cp-definition} \\
    &\sim \frac{3}{16\pi Dt } \exp\left[ -\frac{1}{4Dt}\vec{r}_{pp'}^2 \right] \label{eqn:Cp-analyitcal}
\end{align}
with $\vec{r}_{pp'} \equiv \vec{r}_p - \vec{r}_{p'}$ the difference between the positions of the centers of the plaquettes $p$ and $p'$.
The spatial profile from the automaton circuit is contrasted with the Gaussian prediction~\eqref{eqn:Cp-analyitcal} in Figs.~\ref{fig:crossing-diffusion}\hyperref[fig:crossing-diffusion]{(b)} and \hyperref[fig:crossing-diffusion]{(c)} for $L=256$.
In Fig.~\ref{fig:crossing-diffusion}\hyperref[fig:crossing-diffusion]{(b)}, we plot the two-dimensional correlation function as a function of position at a fixed time satisfying $Dt \ll L^2$, while Fig.~\ref{fig:crossing-diffusion}\hyperref[fig:crossing-diffusion]{(c)} shows a scaling collapse of the full profile for various times [all of which satisfy the condition $t \gtrsim 10$ identified in Fig.~\ref{fig:crossing-diffusion}\hyperref[fig:crossing-diffusion]{(a)}].
Figure \ref{fig:crossing-diffusion} therefore confirms that domain wall crossings diffuse in \emph{typical} sectors, if the $\CZ_p=1$ constraint can be violated.


\subsection{\texorpdfstring{$\CZ_p=1$}{}: Dynamics of closed loops}

Next, we look at correlation functions where the limit $J \to \infty$ is taken before the limit of infinite temperature.
In this case, we work in the constrained Hilbert space with no domain wall crossings, i.e., $\CZ_p = 1$ for all plaquettes $p$.
As discussed in the main text, in addition to the $\sim 2^L$ dynamically frozen scar states, the constrained Hilbert space also possesses additional sectors 
associated with different numbers and orientations of winding domain wall loops. The scar states identified in the main text correspond to a dense packing of these winding loops, forbidding them from fluctuating. When the packing ceases to be dense, the winding loops are able to move under local dynamics while satisfying the no-crossing constraint. Additionally, contractible loops fluctuate in and out of existence in the interstitial vacua separating the winding loops. However, the number of winding loops is unable to change under local dynamics.


\subsubsection{Hydrodynamics of closed loops}

Since $\rho_i(\vec{r})$ form closed loops, we expect that the late time and long wavelength relaxation of the system should be described by a hydrodynamic theory with a one-form symmetry.
Explicitly, to lowest order in derivatives, (a coarse-grained variant of) the vector-valued density $\rho_i$ should evolve according to
\begin{equation}
    \partial_t \rho_i + \epsilon_{ij} \partial_j J = 0
    \, ,
    \label{eqn:one-form-hydro}
\end{equation}
for a scalar current $J(\rho)$. The presence of the two-dimensional Levi-Civita symbol, $\epsilon_{ij}$, implies that the theory conserves an infinite family of charges parametrized by scalar functions~\cite{fractonMHD}, which we show below.
A vector-valued function $f_i(\vec{r})$ defines a conserved charge $Q[f]$ if it satisfies
\begin{equation}
    \dot{Q}[f] = \frac{\mathrm{d}}{\mathrm{d}t} \int \mathrm{d}^2\vec{r} \, f_i(\vec{r}) \rho_i(\vec{r}) = 0
    \, .
    \label{eqn:conserved-f}
\end{equation}
For a $\rho_i$ that evolves in time according to Eq.~\eqref{eqn:one-form-hydro}, we can integrate Eq.~\eqref{eqn:conserved-f} by parts to obtain
\begin{equation}
    \int \mathrm{d}^2\vec{r} \, J \epsilon_{ij} \partial_j f_i(\vec{r}) \stackrel{!}{=} 0
    \, ,
\end{equation}
in order for $f_i(\vec{r})$ to correspond to a conserved quantity. Commutativity of partial derivatives and antisymmetry of $\epsilon_{ij}$ implies that an $f_i(\vec{r})$ obeying
\begin{equation}
    f_i(\vec{r}) = \partial_i \Phi(\vec{r})
    \, ,
\end{equation}
will define a conserved charge for each scalar function $\Phi(\vec{r})$. The nature of the corresponding conserved charges can be elucidated by choosing a basis of indicator functions of the form
\begin{equation}
    \Phi_V(\vec{r}) =
    \begin{cases}
        1 &\text{ for } \vec{r} \in V , \\
        0 &\text{ otherwise,}
    \end{cases}
\end{equation}
for some choice of volume $V$.
With this choice of $\Phi(\vec{r})$, observe that the conserved charges become 
\begin{equation}
    \int \mathrm{d}^2\vec{r} \, \delta(\vec{r} \in \partial V) \hat{n}_i(\vec{r}) \rho_i(\vec{r}) = \int_{\partial V} \rho_i \epsilon_{ij} \mathrm{d}x_j
    \, ,
    \label{eqn:conserved-flux}
\end{equation}
where $\hat{n}_i$ is the unit vector normal to the boundary $\partial V$. Equation~\eqref{eqn:conserved-flux} shows that the flux of $\rho_i$ through (codimension one) surfaces is conserved, as is required for a theory of closed loops.
The simplest constitutive relation between current and the vector-valued charge density is $J = D \epsilon_{ij}\partial_i \rho_j$, leading to a linear hydrodynamic description
\begin{equation}
    \partial_t \rho_i + D \epsilon_{ij}\epsilon_{k\ell} \partial_j \partial_k \rho_\ell
    = 0  .
    \label{eqn:one-form-hydro-expanded}
\end{equation}
It turns out that the rank-four tensor $\epsilon_{ij}\epsilon_{k\ell}$ is the most general tensor structure permitted by $D_4$ symmetry (invariant under $2\pi/4$ rotations and a reflection) and the requirement that $\rho_i$ exhibits a one-form symmetry. In momentum space, this constitutive relation implies that $\rho_i \propto k_i$ cannot decay (since the equations of motion preserve the local constraint $0 = \partial_i \rho_i \leftrightarrow k_i \rho_i$), while $\rho_i$ orthogonal to $k_i$ is a quasinormal mode with $i\omega = D(k_x^2 + k_y^2)$.
For a more thorough discussion of theories analogous to Eq.~\eqref{eqn:one-form-hydro}, and their higher-rank generalizations, we refer the reader to Ref.~\cite{fractonMHD}.

\begin{figure*}[t]
    \centering
    \includegraphics[width=\linewidth]{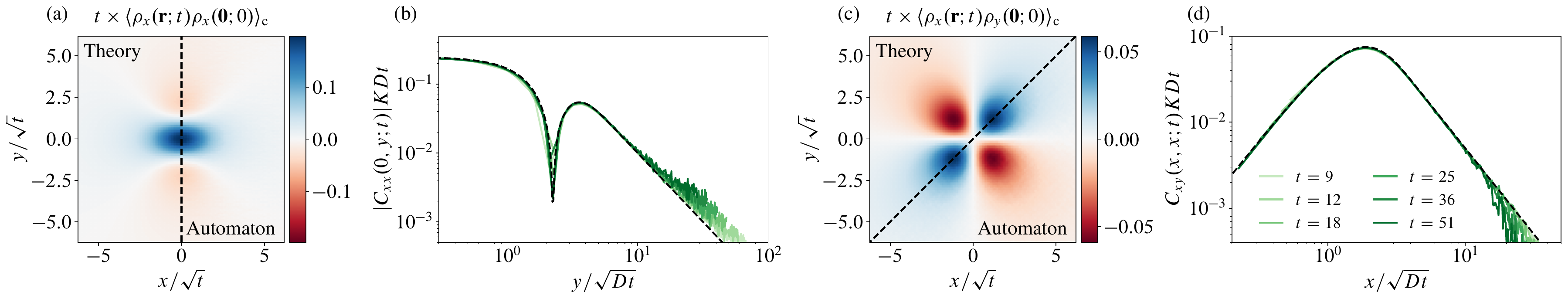}
    \caption{Comparison between correlation functions obtained from hydrodynamics with a one-form symmetry and those obtained from the stochastic automaton circuit. (a) Rescaled two-dimensional profile $C_{xx}(\vec{r}; t)$, with $C_{ij}(\vec{r}; t)\equiv \langle \rho_i(\vec{r}; t) \rho_j(\vec{0}; 0) \rangle$ obtained from the automaton circuit (right panel), compared with the theoretical prediction in Eq.~\eqref{eqn:Cxx-unequal-time} (left panel). (b) Scaling collapse of the correlation function $C_{xx}(\vec{r}; t)$ along the line $x=0$ for various times [indicated in the legend of (d)]. (c) Rescaled two-dimensional profile $C_{xy}(\vec{r}; t)$ obtained from the automaton circuit (lower right panel), compared with the theoretical prediction in Eq.~\eqref{eqn:Cxy-unequal-time} (upper left panel). (d) Scaling collapse of the correlation function $C_{xy}(\vec{r}; t)$ along the line $x=y$ for various times.}
    \label{fig:one-form-automaton}
\end{figure*}


\subsubsection{Infinite temperature dynamics of loops}

In the largest sector, which contains no winding loops ($\mathcal{Q}_x = \mathcal{Q}_y = 0$), the entropy of a given configuration of the system can be accounted for by a Gaussian weight functional $P[\rho_i(\vec{r})] \propto \exp(-S[\rho_i(\vec{r})])$, as in models exhibiting a ``Coulomb phase''~\cite{HenleyAnnuRev}, with
\begin{equation}
    S[\rho_i(\vec{r})]  = \frac{K}{2} \int \mathrm{d}^2\vec{r} \, (\epsilon_{ij} \partial_j h)^2 = \frac{K}{2} \int \mathrm{d}^2\vec{r}\, (\nabla h)^2
    \, .
    \label{eqn:gaussian-weight}
\end{equation}
The scalar field $h(\vec{r})$ is interpreted as a height field in the ``rough'' phase~\cite{HenleyAnnuRev} corresponding to the coarse-grained field $\rho_i(\vec{r}) = \epsilon_{ij} \partial_j h$.
Physically, typical (infinite temperature) states belonging to the $\mathcal{Q}_x = \mathcal{Q}_y = 0$ sector will have lots of short loops, which coarse grain to small values of $\rho_i(\vec{r})$; larger values of $|\rho|$ are entropically suppressed~\cite{HenleyAnnuRev, CastelnovoAnnuRev}. The Gaussian weight~\eqref{eqn:gaussian-weight} directly determines the equal-time correlation functions of the field $\rho_i(\vec{r})$:
\begin{subequations}
\begin{align}
    C_{xx}(\vec{r}; 0) \equiv \langle \rho_x (\vec{r}) \rho_x(0) \rangle &\sim \frac{1}{2\pi K}  \frac{x^2 - y^2}{r^4} \\
    C_{xy}(\vec{r}; 0) \equiv \langle \rho_x (\vec{r}) \rho_y(0) \rangle &\sim \frac{1}{\pi K}  \frac{xy}{r^4} 
\end{align}%
\label{eqn:equal-time-rho}%
\end{subequations}
at large separations $r\equiv | \vec{r} | \gg 1$. The divergence-free constraint gives rise to power-law decaying spatial correlations. At unequal times, the correlation functions are determined by the hydrodynamic equation of motion in Eq.~\eqref{eqn:one-form-hydro-expanded}. 
Using the height model~\eqref{eqn:gaussian-weight}, the correlation functions at unequal times can in fact be evaluated analytically. 
For simplicity of notation, we define dimensionless variables $\bar{x} \equiv x/\sqrt{Dt}$ and similarly for $\bar{y}$ and $\bar{r}$. In terms of these variables, Eq.~\eqref{eqn:equal-time-rho} is generalized to
\begin{subequations}
\begin{align}
    \langle &\rho_x (\vec{r}, t) \rho_x(\vec{0}, 0) \rangle \sim \frac{1}{2\pi K Dt} \times \notag \\ &  \frac{1}{\bar{r}^4} \left[ \bar{x}^2 - \bar{y}^2 + \left( \frac12(\bar{y}^4 +\bar{x}^2\bar{y}^2)  +\bar{y}^2 - \bar{x}^2) \right) e^{-\frac14 \bar{r}^2}  \right] \label{eqn:Cxx-unequal-time} \\
    \langle &\rho_x (\vec{r}, t) \rho_y(\vec{0}, 0) \rangle \sim \frac{1}{\pi K Dt} \times \notag \\ & \qquad\qquad\qquad\qquad  \frac{\bar{x}\bar{y}}{\bar{r}^4} \left[1 - \left( 1 + \frac14 \bar{r}^2 \right) e^{-\frac14 \bar{r}^2}  \right] \label{eqn:Cxy-unequal-time} 
\end{align}%
\label{eqn:C-unequal-time}%
\end{subequations}
which reduce to the equal time correlation functions as $t \to 0^+$. In terms of the dimensionless variables $\bar{x}$ and $\bar{y}$, the long-distance power-law decay of the correlation functions for $\bar{r} \gg 1$ is preserved, but the short-distance divergence is smoothed out at $\bar{r} \ll 1$.
The analytical results in Eq.~\eqref{eqn:C-unequal-time} are compared with the stochastic automaton circuit (sampling only the largest sector) in Fig.~\ref{fig:one-form-automaton}.
The analytical results contain two phenomenological parameters: the diffusion constant $D$ and the stiffness constant $K$, which penalizes spatial fluctuations of the height field $h$. These two parameters are obtained from the scaling collapses in Fig.~\ref{fig:one-form-automaton}, and all plots use the same values of $D$ and $K$.
In all panels, there is excellent agreement between the predictions of hydrodynamics with a one-form symmetry \eqref{eqn:C-unequal-time} and the automaton circuit; the profile spreads diffusively in time, but with long-range, power-law tails, which decay asymptotically as $\sim \bar{r}^{-2}$.


\section{Signatures in exact diagonalization}

In the main text we argued that the ``foliated domain wall'' configurations in systems of finite size are given dynamics at the smallest of (i) $Jt \sim (J/h)^L$, corresponding to perturbative processes that act on non-contractible winding loops of the torus, and (ii) the prethermal time scale, corresponding to $Jt \sim \exp(c n_*)$ where $n_* \sim J/h$ (up to logarithmic corrections) and $c>0$ is an $O(1)$ constant. Qualitatively, the prethermal time scale coincides with the time after which perturbation theory breaks down. 
Here, we verify the robustness of the scar states and their signatures using a combination of automaton circuits and exact diagonalization.

We look at two quantities that give rise to signatures of the scar states' slow dynamics in systems of finite size. The first is the \emph{fidelity} $\mathcal{F}(t)$, which quantifies the overlap between the state at time $t$, $\ket{\psi(t)} = e^{-iHt}\ket{\psi(0)}$, and system's the initial state, $\ket{\psi(0)}$,
\begin{equation}
    \mathcal{F}(t) = |\langle \psi(0) | e^{-i H t} | \psi(0) \rangle|^2
    \, .
    \label{eqn:fidelity-defn}
\end{equation}
The second is a sublattice-magnetization operator, whose sign structure is determined by the initial scar state of interest (alternatively, for product states in the computational basis, it can be viewed as the autocorrelation function $\langle\sigma_i^z(t)\sigma_i^z(0)\rangle$). In contrast to Eq.~\eqref{eqn:fidelity-defn}, the sublattice-magnetization operator corresponds to a sum of local operators, and will turn out to produce signatures of the slow dynamics that are significantly more ``robust'' in the thermodynamic limit, as we argue below. In particular, we evaluate 
\begin{subequations}
\begin{align}
    N \mathcal{O}(t) &= \langle \psi(t) | \sum_{i} s_{i}^{\,} \sigma_{i}^z | \psi(t) \rangle \\
    &= \langle  \psi(t) |  M_0 | \psi(t) \rangle - \langle  \psi(t) |  M_1 | \psi(t) \rangle 
    \, ,
\end{align}%
\label{eqn:sublattice-M}%
\end{subequations}
where the sign $s_{i} = \pm 1$ is equal to $\langle \sigma_{i}^z \rangle$ evaluated in the initial scar state of interest (which, we recall, is a product state in the computational basis $\ket{\sigma_i^z}$). Alternatively, in the second line, we defined $M_{1} = \sum_{i | b_i = 1 } \sigma_i^z$, which is the magnetization of the region of spins that were initially in the `1' state, and $M_0 = \sum_{i | b_i = 0} \sigma_i^z$, which is the magnetization of the complementary region. The normalization of both quantities, $\mathcal{F}(t)$ and $\mathcal{O}(t)$, is such that $\mathcal{F}(0) = \mathcal{O}(0) = 1$ at time $t=0$ when evaluated in the appropriate scar state.


\subsection{Schrieffer-Wolff transformation} 

Before performing any numerical simulations, we first discuss some expectations based on the Schrieffer-Wolff perturbation theory procedure. The effective Hamiltonian that describes time evolution up to the prethermal time scale is related to the original Hamiltonian $H$ via a unitary transformation
\begin{equation}
    H' = e^{i{S}}{H}e^{-i{S}} = {H} + [i{S}, {H}]+ \frac{1}{2!} [i{S}, [i{S}, {H}]]+ \ldots
    \label{eqn:S-W}
\end{equation}
parametrized by the Hermitian operator $S = S^\dagger$.
The operator $S$ is chosen in such a way as to conserve the number of crossings up to a particular order in $h/J$ (by removing any terms generated at the previous order in perturbation theory that do not conserve the number of domain wall crossings).
The eigenstates of the effective Hamiltonian ${H}'$ are then dressed by the operator $e^{-i{S}}$ to obtain the eigenstates of the original Hamiltonian ${H}$.
The ``true'' scar states, i.e., those states that give rise to almost exactly frozen dynamics, are the eigenstates of $H'$~\eqref{eqn:S-W} truncated at a particular order in $h/J$. These states are product states in the computational basis, dressed by $e^{-iS}$, which creates a small but nonvanishing density of crossings.

More precisely, let the state $\ket{\phi}$ be a foliated pattern of domain walls in the computational basis, and suppose that the initial state of the system is its dressed counterpart, $\ket{\psi(0)} = e^{-iS}\ket{\phi}$.
In this case, Eq.~\eqref{eqn:sublattice-M} evaluates to
\begin{subequations}
\begin{align}
    N\mathcal{O}(t) &= \bra{\phi} e^{iS} e^{iHt} (M_1 - M_0) e^{-iHt} e^{-iS} \ket{\phi} \\
    &= \bra{\phi} e^{iH't} (M_1' - M_0') e^{-iH't} \ket{\phi}
\end{align}%
\label{eqn:O-strict-conservation}%
\end{subequations}
where $M_i' = e^{i{S}} M_i e^{-i{S}}$.
Since $\ket{\phi}$ is a trivial eigenstate of $H'$ (up to corrections exponentially small in $J/h$) that exhibits no dynamics, $\mathcal{O}(t)$ does not exhibit any dynamics (up to quantities that are exponentially suppressed in $J/h$). This follows from the fact that each $M_i$ is a sum of local operators. Conversely, suppose that we begin in the state $\ket{\phi}$ \emph{without} the dressing operator. This is the case most relevant to both experiments and numerical simulations, where product states in the real-space basis are most easily prepared.
In this case, 
\begin{equation}
    N \mathcal{O}(t) = \bra{\phi} e^{-iS} e^{iH't} (M_1' - M_0') e^{-iH't} e^{iS} \ket{\phi}
    \, .
    \label{eqn:O-approx-conservation}
\end{equation}
The state $e^{iS} \ket{\phi}$ is \emph{not} an approximate eigenstate of $H'$. Instead, it corresponds to a nontrivial superposition of states with a crossing density of order $\sim h/J$ (since the operator $iS$ creates pairs of domain wall crossings) and Eq.~\eqref{eqn:O-approx-conservation} can therefore exhibit nontrivial dynamics, even at ``short'' times set by $ht =O(1)$. While $\bra{\phi}e^{iS}\ket{\phi}$ becomes vanishingly small as $L \to \infty$, the two states $\ket{\phi}$ and $e^{iS} \ket{\phi}$ differ by a small but nonvanishing density of crossings set by the small control parameter $h/J$, such that they \emph{locally} look almost identical as $h/J \to 0$. We therefore expect that experimentally preparable product states should exhibit signatures of the scar states' slow dynamics, in spite of their vanishingly small overlap in the thermodynamic limit.
However, in the presence of a finite density of crossings, the motion of crossings throughout the system can disrupt the underlying spin configuration, leaving behind a `trail of destruction', since they propagate via a sequence of spin flips that destroy the underlying domain wall pattern. The precise time scales over which a nonzero crossing density can degrade the underlying spin configuration will be discussed in the next section.

The fidelity $\mathcal{F}(t)$ is expected to be significantly less robust, but will still exhibit signatures of the scar states in systems of finite size (which we study using exact diagonalization below).
While the corrections to $H'$ that do not conserve the number of crossings are exponentially small in $J/h$, they are also proportional to system volume $L^2$. This means that, in the thermodynamic limit $L\to \infty$, the effective Hamiltonian truncated at a certain order provides an accurate description of the expectation values of local operators [e.g., $\mathcal{O}(t)$, as in Eq.~\eqref{eqn:O-strict-conservation}], but \emph{not} of the time evolution of individual states, at least in the sense that $\bra{\psi}{U_\text{app}(t)}\ket{\psi} \to 0$ in the thermodynamic limit, where $U_\text{app}$ is the approximate time evolution operator obtained by truncating the effective Hamiltonian $H'$ at a certain order in $h/J$.
Even in systems of finite size and at values of $h/J$ sufficiently small that we are able to accurately describe the time evolution of individual states by truncating the effective Hamiltonian in Eq.~\eqref{eqn:fidelity-defn}, the fidelity can \emph{still} decay if the system is initialized in a scar state $\ket{\phi}$ without its dressing operator, since the small but nonvanishing density of domain wall crossings are able to move around under the dynamics generated by $H'$.


\subsection{Motion of crossings}
\label{sec:crossing-motion}

\begin{figure}
    \centering
    \includegraphics[width=\linewidth]{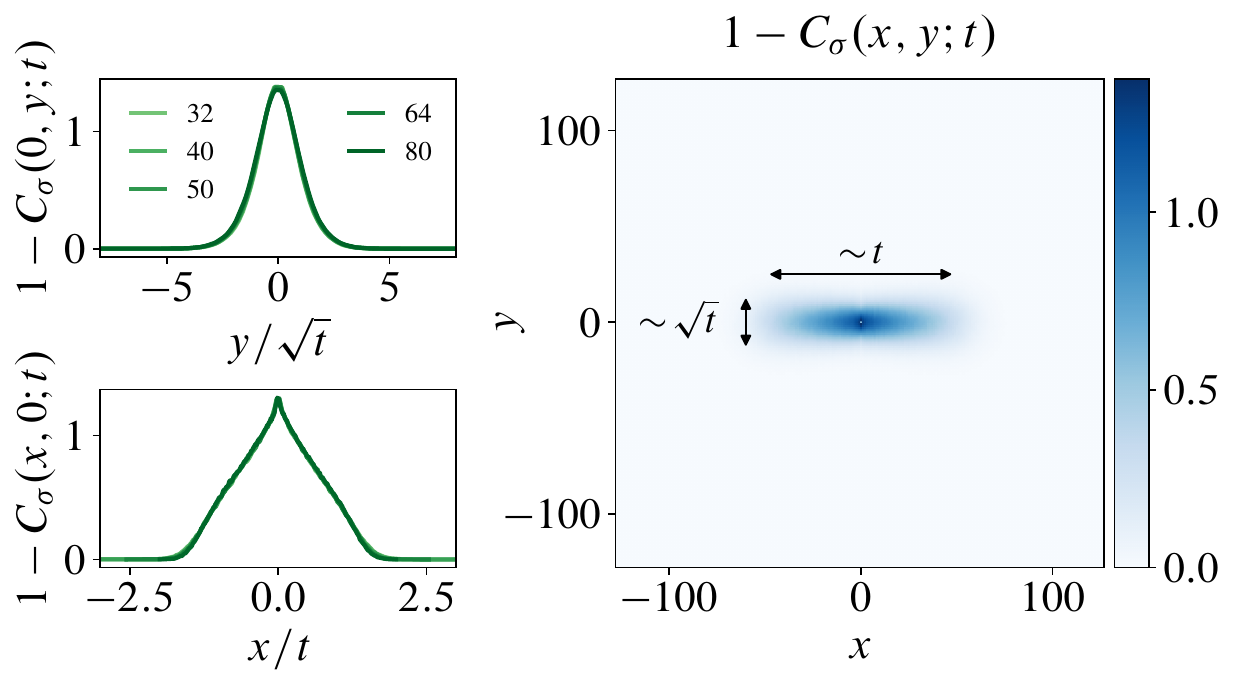}
    \caption{Illustration of the anisotropic spreading of crossings when propagating on top of a scar pattern. A defective spin in a scar state creates a pair of neighboring crossings. Under automaton dynamics, these crossings move along the domain walls at some constant speed (on average), and exhibit diffusive motion in the transverse direction. This behavior is illustrated by the spatially resolved autocorrelation function $C_\sigma(\vec{r}; t) \equiv \langle \sigma_\vec{r}^z(t) \sigma_\vec{r}^z(0) \rangle$, where the system is initialized in a random scar state with a defective spin at the center. The `ballistic' and diffusive spreading in the directions parallel (bottom) and perpendicular (top) to the domain walls is verified in the left panels.}
    \label{fig:anisotropic-spread}
\end{figure}

To discuss quantitatively how the presence of a nonvanishing density of crossings disrupts the underlying spin pattern, we first consider the motion of an isolated crossing atop a foliated pattern of domain walls. In order to access large system sizes, we will restrict our attention here to dynamics generated by stochastic cellular automaton circuits~\footnote{We caution that quantum circuit dynamics may not accurately capture the scaling of diffusion constants in the Hamiltonian system at low defect density. See, for instance, Ref.~\cite{Chen2020butterfly}.}. In this context, we discussed in Sec.~\ref{sec:crossing-diffusion} how the density of crossings asymptotically exhibits isotropic diffusion at infinite temperature. In contrast, for a low density of crossings decorating a low-entropy scar state, the dynamics is affected by the local direction of the domain wall loops: There exists an entropic bias that leads to a nonzero average velocity in the direction parallel to the domain walls. This can be illustrated, for example, by the configuration:
\begin{equation*}
    \includegraphics[width=0.2\linewidth]{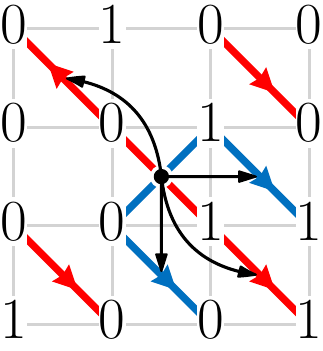}
\end{equation*}
which represents part of a larger system (in the presence of periodic boundaries, crossings must come in pairs). The black arrows represent the motion of the crossing under the permitted (single) spin flips, which preserve the number of domain wall crossings, hopping the crossing to an adjacent plaquette.
Three of the four possible spin flips have a positive projection onto the southwest direction, while just one hop has a negative projection.
In the automaton circuits, this leads to a biased random walk of the crossing parallel to the domain walls. In the direction transverse to the domain walls (northeast, say), there is an equal probability of hopping with positive or negative projection. Consequently, the crossings exhibit diffusive spreading in the transverse direction. Since the crossings propagate via a sequence of spin flips, they degrade the original scar pattern as they move throughout the lattice. Additionally, once the scar pattern has been disrupted, spins in the vicinity of the crossings' trails may also become flippable.

The above arguments are verified numerically using the stochastic automaton circuits introduced in Sec.~\ref{sec:automaton}. We initialize the system in a random scar state with maximal $|\mathcal{Q}_x|$ (obtained by shifting the repeated pattern `0011' in adjacent columns either to the left or to the right with equal probability), and flip a single spin at the center of the system, thereby creating a neighboring pair of crossings with opposite `charge' (i.e., one acts as a source of $\Div\boldsymbol{\rho}$, and the other as a sink).
The subsequent dynamics of the system is studied by following the evolution of the spatially resolved autocorrelation function $C_\sigma(\vec{r}; t) \equiv \langle \sigma_\vec{r}^z(t) \sigma_\vec{r}^z(0)  \rangle$, where the average $\langle \,\cdots\, \rangle$ is over independent realizations of the automaton circuit and over random initial scar states.
Note that, since the initial states explicitly break translation invariance, $C_\sigma(\vec{r}; t)$ is not independent of position $\vec{r}$.
Results are shown in Fig.~\ref{fig:anisotropic-spread}. Because the ensemble of scars from which we sample maximizes $|\mathcal{Q}_x|$, the crossings exhibit a biased random walk along $x$~\footnote{Note that the random scar states are not perfectly aligned along $x$; the typical value of $|\mathcal{Q}_y|$ is set by $\sqrt{L}$. This leads to an additional source of diffusive broadening in the transverse direction.}, and diffuse in the transverse ($y$) direction. This leads to the anisotropic degradation of the scar pattern depicted in the right panel of Fig.~\ref{fig:anisotropic-spread}: There exists a region of width $\sim t$ and height $\sim \sqrt{t}$ that spreads from the initial location of the flipped spin, within which the scar pattern is disarranged. The dynamical exponents in the two directions are substantiated in the scaling collapses in the left panels of Fig.~\ref{fig:anisotropic-spread}.

\begin{figure}
    \centering
    \includegraphics[width=0.81\linewidth]{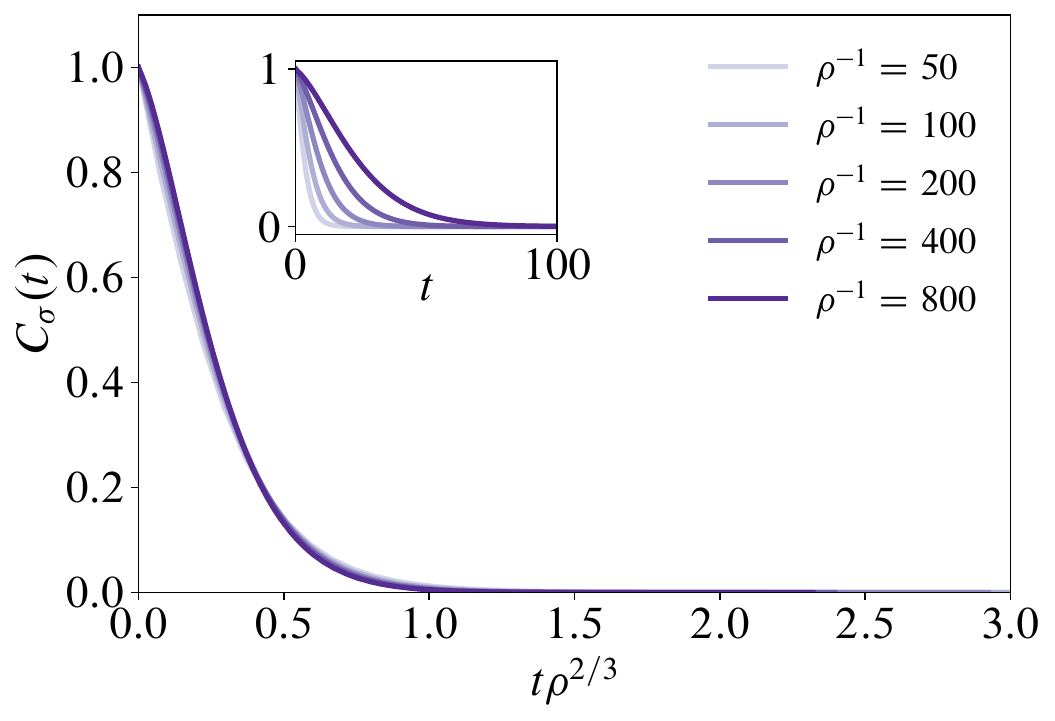}
    \caption{Spatially averaged spin autocorrelation function $C_\sigma(t)$ for random scar states degraded by various densities $\rho$ of defective spins, which create neighboring pairs of crossings. The raw data (shown in the inset) exhibit a collapse upon rescaling time by $t \to t  \rho^{2/3}$, which reflects the fact that crossings spread anisotropically when propagating atop a foliated pattern of domain walls, as depicted in Fig.~\ref{fig:anisotropic-spread}. The data are obtained using a system of size $L=512$ and averaged over $2^{13}$ histories.}
    \label{fig:defective-scar}
\end{figure}

\begin{figure*}[t]
    \centering
    \includegraphics[width=\linewidth]{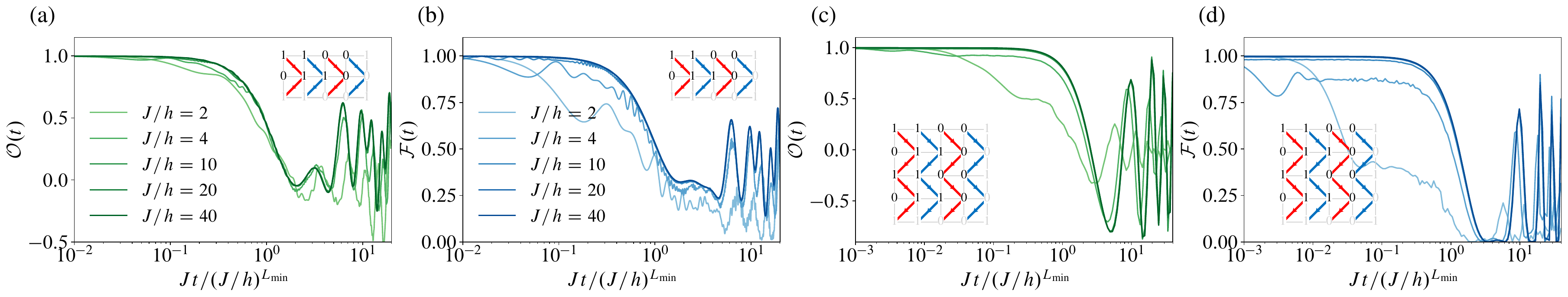}%
    \caption{(a), (c): Expectation value of the sublattice-magnetization operator $\mathcal{O}(t)$ with a scar state $\ket{\sigma_i^z}$ in the computational basis for $(L_x, L_y) = (4, 2)$ [(a)] and $(4, 4)$ [(c)]. The data exhibit an approximate collapse upon rescaling time by $(J/h)^{L_\text{min}}$, with $L_\text{min}=\min(L_x, L_y)$, which corresponds to the time scale at which perturbative processes that wrap around the torus are permitted. (b), (d): the fidelity $\mathcal{F}(t)$ evaluated for the same initial states. The quality of the collapse is poorer, with the fluctuations of $\mathcal{F}(t)$ from unity at times $Jt \lesssim (J/h)^{L_\text{min}}$ larger than those observed for $\mathcal{O}(t)$ in (a), (c). Panels (c) and (d) share legends with (a) and (b), respectively.}
    \label{fig:exact-diag}
\end{figure*}

Armed with the knowledge of how crossings propagate throughout the system on top of typical scars, we can resolve the question of how a finite density of defects, $\rho \sim h/J$, affects the scar states' dynamics.
If the motion of the crossings led to diffusive spreading of $C_\sigma(\vec{r}; t)$, we would expect that the scar pattern would be fully degraded after a time $t \sim \ell^2 \sim \rho^{-1}$, where $\ell = (J/h)^{1/d}$ is the typical separation between defects in the initial state.
However, the anisotropic spreading of the $C_\sigma(\vec{r}; t)$ leads to a different time scale. Intuitively, this time scale, which corresponds to the time at which the anisotropic regions in Fig.~\ref{fig:anisotropic-spread} intersect, can be understood as the time at which the area of the region of degradation ($t \times \sqrt{t}$) equals the characteristic area associated with each defective spin, i.e., $t \sqrt{t} \sim \ell^2$, giving rise to a characteristic time scale $t \sim \ell^{4/3} \sim \rho^{-2/3}$.
This scaling is demonstrated in Fig.~\ref{fig:defective-scar}, where random scar states are degraded by varying densities $\rho$ of flipped spins. The autocorrelation function averaged over the full system, $C_\sigma(t) = L^{-2} \sum_\vec{r} C_\sigma(\vec{r})$, is measured as a function of time and exhibits a collapse when plotted as a function of $t\rho^{2/3}$.  
Consequently, the time scale for $C_\sigma(t)$ to decay by an appreciable fraction is enhanced by a factor $(J/h)^{2/3} \gg 1$ for a state that is ``close'' to a scar state relative to a `typical' (infinite temperature) initial product state.

We conclude by noting that the above discussion relates only to the experimental preparability of the scar states, which will only ever be approximate.
We have shown that product states, despite having vanishing overlap with the scars in the thermodynamic limit, exhibit signatures of the scar states' slow dynamics.
In principle, the time scale $ht \sim (J/h)^{2/3}$ identified herein can be parametrically enhanced by reducing the density of defective spins in the initial state in the Schieffer-Wolff-transformed basis. The exact scars correspond to a limiting case of this procedure with vanishing defect density.


\subsection{Numerical results} 

The above expectations are borne out in Fig.~\ref{fig:exact-diag}. We perform exact diagonalization on the two smallest systems that are compatible with the foliated domain wall pattern discussed in the main text. Namely, we consider $(L_x, L_y) = (4, 2)$ and $(4, 4)$. The former only hosts $8$ scars corresponding to domain walls that wrap around the torus parallel to $y$ (the particular scar state is shown in the inset of each figure), while the latter, being square, exhibits all of the $2^{L+2}-8=56$ scar states discussed in the main text.
For both system sizes, the operator $\mathcal{O}(t)$ exhibits dynamics at a time scale set by $Jt \sim (J/h)^{L_\text{min}}$ for sufficiently small values of $h/J \lesssim L^{-2}$, as predicted by perturbation theory, where $L_\text{min} = \min(L_x, L_y)$. That is, for $h/J \lesssim L^{-2}$ there are effectively no defects in the Schrieffer-Wolff-transformed basis and the product state behaves like the true scar state. Put differently, the scar state and the corresponding product state have appreciable, $O(1)$, overlap, leading to a diverging time scale $Jt \sim (J/h)^{L_\text{min}}$ that is also witnessed by the fidelity.
For larger values of $h/J$, the operator $\mathcal{O}(t)$ and the fidelity $\mathcal{F}(t)$ exhibit significant deviations from unity prior to the time scale $Jt \sim (J/h)^{L_\text{min}}$, since the initial state contains a nonvanishing number of crossings (on average) in the Schrieffer-Wolff-transformed basis, which can subsequently propagate throughout the system, as described in Sec.~\ref{sec:crossing-motion}. Intriguingly, even in the regime $h/J \gtrsim L^{-2}$, both $\mathcal{O}(t)$ and $\mathcal{F}(t)$ exhibit nonzero plateaux, suggesting that nonvanshing overlap with the subspace of scars remains in this regime (although this is likely a finite size effect).

\end{document}